\documentclass[iop]{emulateapj}

\usepackage{graphics}
\usepackage{epsfig}
\usepackage{graphicx}
\usepackage{color}
\usepackage{pstricks}
\usepackage{hyperref}
\usepackage{afterpage}
\usepackage{epstopdf}
\usepackage{epsfig}
\usepackage{natbib}

\renewcommand{\thefigure}{\arabic{figure}}

\newcommand{\etal}{et~al.\ }

\newcommand{\magsec}{mag/arcsec$^2$}

\newcommand{\LCDM}{$\Lambda$CDM}

\newcommand{\msun}{M_{\odot}}

\def\kms{km~s$^{-1}$}

\def\msun{$M_\odot$}

\shorttitle{The intriguing HI1232+20 system}
\shortauthors{Janowiecki \etal}

\received{23 Dec 2014}
\revised{3 Feb 2015}

\begin{document}

\title{(Almost) Dark HI Sources in the ALFALFA Survey: The
        Intriguing Case of HI1232+20}

\author{Steven Janowiecki\altaffilmark{1},
Lukas Leisman\altaffilmark{2},
Gyula J\'ozsa\altaffilmark{3,4,5},
John J. Salzer\altaffilmark{1},
Martha P. Haynes\altaffilmark{2},
Riccardo Giovanelli\altaffilmark{2},
Katherine L. Rhode\altaffilmark{1},
John M. Cannon\altaffilmark{6},
Elizabeth A. K. Adams\altaffilmark{7},
William F. Janesh\altaffilmark{1}
}
\email{sjanowie@astro.indiana.edu
}

\altaffiltext{1}{Department of Astronomy, Indiana University, 727 East
  Third Street, Bloomington, IN 47405, USA}

\altaffiltext{2}{Center for Radiophysics and Space Research, Space
  Sciences Building, Cornell University, Ithaca, NY 14853, USA}

\altaffiltext{3}{SKA South Africa, Radio Astronomy Research Group, 3rd
  Floor, The Park, Park Road, Pinelands, 7405, South Africa}

\altaffiltext{4}{Rhodes University, Department of Physics and
  Electronics, Rhodes Centre for Radio Astronomy Techniques \&
  Technologies, PO Box 94, Grahamstown, 6140, South Africa}

\altaffiltext{5}{Argelander-Institut f\"ur Astronomie, Auf dem H\"ugel
  71, 53121 Bonn, Germany}

\altaffiltext{6}{Department of Physics \& Astronomy, Macalester
  College, 1600 Grand Avenue, Saint Paul, MN 55105}

\altaffiltext{7}{Netherlands Institute for Radio Astronomy (ASTRON),
  Postbus 2, 7990 AA, Dwingeloo, The Netherlands}

\slugcomment{revised version 1.6, 3 Feb 2015}

\begin{abstract}



{We report the discovery and follow-up observations of a system
  of three  objects identified by the ALFALFA extragalactic HI survey,
cataloged as (almost) dark extragalactic sources,
i.e., extragalactic HI detections with no discernible counterpart in
publicly available, wide-field, imaging surveys.}
We have obtained deep optical imaging with WIYN pODI and HI
synthesis maps with WSRT of {the HI1232+20 system}. The source
with the highest HI flux has a newly discovered ultra-low surface
brightness (LSB) optical counterpart associated with it, while the
other two sources have no detected optical counterparts in our
images. 
Our optical observations show that the detected LSB optical 
counterpart has a peak 
surface brightness of $\sim$$ 26.4$~\magsec \, in $g'$, which is
exceptionally faint. This source (AGC~229385) has the largest
{accurately measured}
HI mass-to-light ratio of an isolated object:
$M_{HI}/L_{g'}$$=$$46$~$M_\odot/L_\odot$,
{and has an HI mass of $7.2$$\times$$10^8 M_\odot$.}
The other two HI sources {(with HI masses $2.0$$\times$$10^8$ and
  $1.2$$\times$$10^8 M_\odot$)} without optical counterparts have upper
limit surface brightnesses of $27.9$ and $27.8$~\magsec \, in $g'$,
and lower limits on their gas mass-to-light ratio of  $M_{HI}/L_{g'}$
$>$$57$ and $>$$31$~$M_\odot/L_\odot$.
{This system lies relatively close in projection to the Virgo
  Cluster, but velocity flow models indicate that it is located at
  $25$~Mpc, substantially beyond Virgo. The system appears to be quite
isolated, with no known object closer than $\sim$$500$~kpc.}
These HI sources may represent both sides of the threshold
between ``dark'' star-less galaxies and galaxies with stellar
populations. 
We discuss a variety of
possible formation scenarios for the HI1232+20 system.

\end{abstract}

\keywords{}

\section{Introduction}

Low surface brightness (LSB) galaxies are
difficult to detect optically, {and thus may be
  underrepresented in most optically-selected samples used in studies
  of galaxy formation and evolution and their} hierarchical assembly history
(e.g., 
\citealt{williams96},
\citealt{madau98},
\citealt{mcgaugh2000}, 
\citealt{brinchmann04},
\citealt{hopkins06}). 
Some of the LSB galaxies might be those in which star formation has been a
slow and gradual process (\citealt{mcgaugh97},
\citealt{schombert14}) {and some may} provide a source of fresh
gas infall to larger galaxies {through merger and interactions}
\citep{sancisi90}. By 
missing the LSB galaxies in most surveys, we may be 
missing an entire population of galaxies and/or a phase of galaxy
evolution.

{ 
At the extreme end of the LSB galaxy spectrum, \citet{disney76}
predicted the existence of entirely ``dark galaxies'', with no
observable optical stellar counterparts because their surface
brightness is too low. A category of ``crouching giants'', exemplified
by the highly luminous and massive LSB spiral Malin I
(e.g., \citet{lelli2010}), has been identified, but they are 
quite rare. Overall, no large population of unseen LSB
objects has been detected at any wavelength.
}

{
LSB galaxies typically possess substantial reservoirs of atomic
hydrogen, so blind 21-cm surveys represent the best opportunity to
find large populations of the most extreme LSBs.
Two major blind HI surveys, HIPASS
(HI Parkes All Sky Survey, \citealt{doyle05}) and ALFALFA (Arecibo
Legacy Fast ALFA, \citealt{haynes11}), have reached the
conclusion that there is not a significant population of gas-bearing
but optically dark systems. At the same time, there are a number of 
intriguing, unexplained objects detected clearly in HI, showing signs
of ordered motion and coincident with no discernible stellar
counterpart
The best example of such a ``dark galaxy'' remains the
southwestern component of the HI1225+01 system \citep{chengalur95,
  matsuoka12}, although it is important
to note its presence in a common envelope with a visible star
forming dwarf companion.
}

\begin{deluxetable*}{lccccccc}
\tabletypesize{\footnotesize}
\tablewidth{0pt}
\tablecaption{Observed HI Parameters of the HI1232+20 System
\label{HIdat} 
}
\tablehead{ 
\colhead{AGC} &\colhead{Position}  &\colhead{F$_{\rm HI, ALFALFA}$}
&\colhead{F$_{\rm HI}$} &\colhead{V$_{50}$} &\colhead{W$_{50}$}
&\colhead{R$_{\rm HI}$} &\colhead{R$_{5\times10^{19}}$}  \\ 
              &\colhead{J2000}    &\colhead{[Jy~\kms]}    &\colhead{[Jy~\kms]} &\colhead{[\kms]}   &\colhead{[\kms]}   &\colhead{[$'$]} &\colhead{[$'$]} \\ 
\colhead{(1)} &\colhead{(2)}      &\colhead{(3)}          &\colhead{(4)}      &\colhead{(5)}      &\colhead{(6)}          &\colhead{(7)}     &\colhead{(8)} 
}
\startdata
229385  & 12:32:10.3 $+$20:25:24  & $4.87\pm0.04$ & $4.84\pm0.04$ &  $1348\pm1$ & \ $34\pm1$ &  $1.60 \times 0.72$  & $1.87 \times 1.10$  \\
229384  & 12:31:36.4 $+$20:20:06  & $1.36\pm0.03$ & $1.25\pm0.04$ &  $1309\pm1$ & \ $27\pm1$ &  $0.73 \times 0.55$  & $1.00 \times 0.70$   \\
229383  & 12:30:55.3 $+$20:34:04  & $0.81\pm0.06$ & $0.42\pm0.05$ &  $1282\pm4$ & \ $59\pm8$ &  \nodata             & $0.80 \times 0.28$ 
\enddata
\tablecomments{ Observed properties of objects in HI1232+20.
(1) Catalog ID in the Arecibo General Catalog (an internal database
     maintained by M.P.H. and R.G.)
(2) HI centroid position from WSRT
(3) Total integrated HI line flux density measured from ALFALFA data
(4) Total integrated HI line flux density measured from WSRT data
(5) Heliocentric velocity, measured at the 50\% flux level
(6) HI velocity width, measured at the 50\% flux level 
(7) HI radius at a HI surface density of 1 \msun~pc$^{-2}$
  (corresponding to an HI column density of
     1.25$\times$10$^{20}$~cm$^{-2}$), in arcminutes, measured from
     the moment 0 maps assuming 
  a beam of $13''$$\times$$39''$. Uncertainties on all radius measurements
  are $\pm0.06'$. At a distance of 25~Mpc, $1'$ subtends a distance of
  $7$~kpc.
  Note that AGC~229384 is separated into two peaks
  at the 1\msun~pc$^{-2}$ level, and that AGC~229383 never reaches a
  surface density of 1\msun~pc$^{-2}$. Also note that the measurement
  for AGC~229383 only reflects the radius of the NW clump; the SE
  clump reaches a column density of 5$\times$10$^{19}$ over an area of
  $14''$$\times$$7''$.
(8) HI semimajor and semiminor axis at a column density of
  5$\times$10$^{19}$ cm$^{-2}$, in arcminutes, assuming a beam
  of $13''$$\times$$39''$.
}
\end{deluxetable*}

While {some} simulations can produce dark galaxies in the form
of stable gas disks that never {produce} stars \citep{verde02},
others find that star-less galaxies cannot exist for very long before
becoming unstable to star formation \citep{TaylorWebster2005}. 
The presence of HI in some LSB galaxies provides
a key dynamical tracer of the mass in these {extreme}
systems (\citealt{geha06}, \citealt{huang12dwarfs}).
{Detailed kinematic studies are being undertaken to
  study the effects of outflows and feedback in lower mass galaxies,
  \citep{vaneymeren2009}, in order to understand star formation modes
  in these shallow potential wells and low density galaxies.}
Groups have worked to develop models that can simultaneously
explain galaxy scaling relations in the full cosmological context
(e.g., \citealt{dutton07}). 


Recently,
{the ALFALFA} survey has made significant
improvements in the sensitivity and depth of available wide-field
blind HI surveys. ALFALFA has measured 25,000+ HI
sources over 7000 square degrees in a cosmologically significant
volume (\citealt{giovanelli05}, \citealt{haynes11}, Jones \etal in
prep.). ALFALFA has characterized the population of normal galaxies
\citep{huang12}, low mass galaxies \citep{huang12dwarfs}, as well as
probing the HI mass function to lower HI masses than ever before
\citep{martin10}. 

As discussed in \citet{cannon15}, the ALFALFA (Almost) Dark Galaxy
Project has been studying the very small fraction ($\sim$$0.4$\%) of
HI~sources which lack obvious optical counterparts and are isolated
from other sources. Followup observations are ongoing  
and include deep optical imaging and HI synthesis maps. 
{Many objects} turn out to be tidal {in origin, but some
 have very low surface brightness stellar populations at or below the
 detection limits of current wide field imaging surveys.}

In this work we study the newly discovered HI1232+20 system of three
(almost) dark 
{extragalactic HI sources} 
which {were not detected in optical surveys, and are at least
  an order of magnitude less luminous than previously studied LSB
  galaxy populations (e.g., \citealt{schombert11}).}
This paper is organized as follows.
In Section~\ref{disco} we describe the discovery and observations of
this system, and in Section~\ref{results} we show the
results of those observations.
Throughout this work we use a flow model distance \citep{masters05} of
$D$$=$$25$~Mpc to the HI1232+20 system, and we discuss the effects of
distance uncertainty on our conclusions in Section~\ref{denv}. 
In Section~\ref{discussion} we discuss the implications of these objects and
what they might mean in the context of (almost) dark galaxies,
and in terms of extending scaling relationships from normal galaxies.
Section~\ref{summary} contains a brief summary of our
main results. Throughout this work we assume a \LCDM \, cosmology, with
$\Omega_{\rm m}$$=$$0.3$, $\Omega_{\rm \Lambda}$$=$$0.7$, and
$H_0$$=$$70$~\kms~Mpc$^{-1}$.



\begin{figure*}[htb]
\centering
\includegraphics[width=16.5cm]{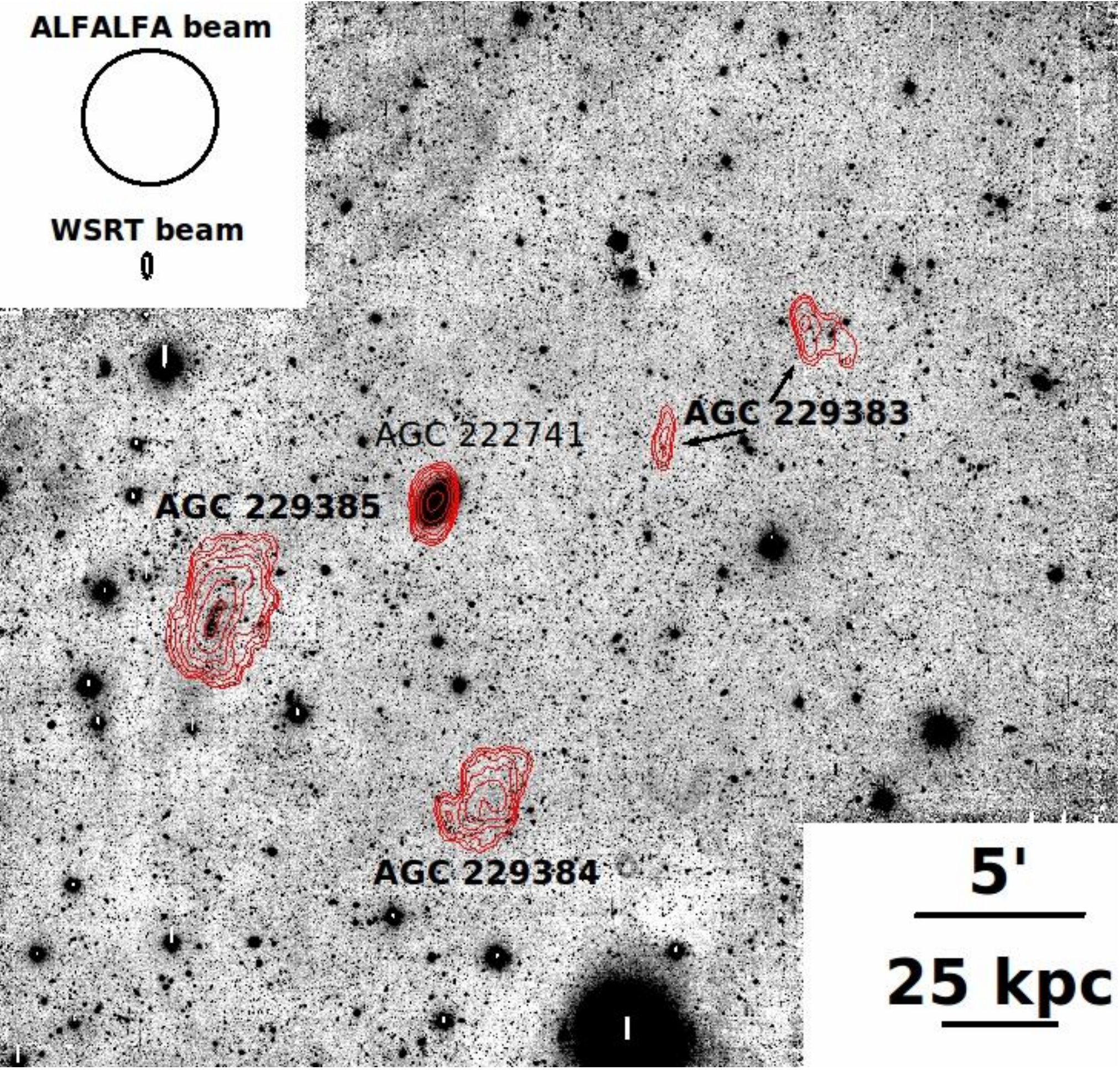} 
\caption{Detection image made from all pODI observations with ALFALFA
  HI sources in the HI1232+20 system labeled and WSRT contours
  overlaid in red. The size of the ALFALFA and WSRT beams are
  indicated by the shapes at the top left corner. The WSRT pointings
  cover nearly the entire area region shown in this image. The
  optical image is binned $12 \times 12$ pixels so that the resulting
  pixels are $1.3''$ on a side, and 
  is aligned north up east left. The WSRT contours show column
  densities spaced logarithmically between 1 and
  64$\times$$10^{19}$~cm$^{-2}$. { The tight packing of the
    lowest contours is 
  likely a result of sigma-clipping, and not a sharp edge in the HI
  distribution. See Section \ref{obs.hi} for more details and about
  sensitivity differences across the image. }
  The spiral galaxy AGC~222741 has a recession velocity of
  $cz$$=$$1884$~km/s, so is significantly more distant than the
  objects in the HI1232+20 system, which have recession velocities
  around $cz$$\sim$$1300$~km/s (see Section~\ref{denv} for details). 
  The WSRT contours of AGC~229383 show that it is separated into two
  components at this sensitivity level, as is discussed further
  in Section \ref{AGC229383}.
  Also visible at the top left of the image is diffuse
  filamentary emission
  from Galactic cirrus, which coincides with features seen in
  far-infrared and ultraviolet images of the same area.\label{allim}
}
\end{figure*}

\section{Observations}\label{disco}

\subsection{ALFALFA discovery of HI1232+20 system}

The ALFALFA survey employs a two-pass, fixed azimuth drift scan
strategy, the details of {which} are described
in previous papers (\citealt{giovanelli05},
\citealt{saintonge07}, \citealt{martin09}, \citealt{haynes11}).
All data are flagged for radio frequency interference (RFI)
interactively, and each grid is examined 
by hand to confirm and improve on sources detected via the automated
methods of \citet{saintonge07}; final source parameters are measured and
cataloged interactively.

Among the ALFALFA (almost) dark 
{extragalactic sources,} 
the HI1232+20
system (comprised 
of sources AGC~229383, AGC~229384, and AGC~229385) was
found to be of particular interest. These three objects are
near each other on the sky and also have similar recession
velocities, so are likely associated with each other. From the ALFALFA 
observations it was clear that these three sources
have significant amounts of gas present, even though 
they do not have readily identifiable optical (stellar) counterparts
in existing optical databases (SDSS, DSS). While they appear on the
sky near AGC~222741 (CGCG~129-006), there is a significant separation
in velocity 
between the sources. AGC~222741 has an HI recession velocity of
$1884$ km/s while the three sources in this sample have
recession velocities of $\sim$$1300$ km/s.

An overlapping archival ultraviolet (UV) image from GALEX GR7 
(Galaxy Evolution Explorer, \citealt{martin05}, \citealt{morrissey07},
Data Release 7, \citealt{bianchi14}) shows a faint diffuse UV source at
the coordinates of AGC~229385 (see Section \ref{galex} for more
details). There is also a hint of a faint 
object at the same position in the DSS2-B image (Digitized Sky
Survey\footnote{The Digitized Sky Surveys were
  produced at the Space Telescope Science Institute under
  U.S. Government grant NAG W-2166. The images of these surveys are
  based on photographic data obtained using the Oschin Schmidt
  Telescope on Palomar Mountain and the UK Schmidt Telescope. The
  plates were processed into the present compressed digital form with
  the permission of these institutions.})
but no source visible at that position in images from SDSS DR9 (Sloan
Digital Sky Survey, Data Release 9, \citealt{ahn12}). Tables
\ref{HIdat} and \ref{abs} contain complete information about the
HI1232+20 system. No optical sources were evident at the locations of
the other two HI detections.

Given the curious nature of this system, we have carried out further
observations to study it in more detail. We have obtained 
deep optical images to 
look for possible faint stellar populations in the sources, and
sensitive HI synthesis observations to resolve the gas
distribution and kinematics in more detail.

\vspace{0.5cm}

\subsection{Deep optical imaging with WIYN pODI}
\label{podi}

The HI1232+20 system was observed with the WIYN\footnote{The WIYN
  Observatory is a joint facility of the
  University of Wisconsin-Madison, Indiana University, the University
  of Missouri, and the National Optical Astronomy Observatory.} 3.5-m
telescope at Kitt Peak National Observatory\footnote{Kitt Peak
  National Observatory, National Optical 
  Astronomy Observatory, which is operated by the Association of
  Universities for Research in Astronomy (AURA) under cooperative
  agreement with the National Science Foundation.} using the partially
populated One Degree Imager (pODI).
Currently,
pODI is made up of 13 Orthogonal Transfer Arrays (OTAs), each of which
is made of sixty-four 480x496 pixel cells.
The OTAs are arranged on the focal plane such that the central 3x3
OTAs cover an area of $24' \times 24'$ with pixels that are $0.11''$ on a
side. The numerous gaps between cells and OTAs require a series of
offset dithered exposures to produce a well-sampled image. Four of the
standard Sloan Digital Sky Survey (SDSS, \citet{gunn98},
\citet{doi10}) $g'$,$r'$,$i'$, and $z'$ filters are available, and
stars from the SDSS catalog photometry are used for standard
photometric calibrations.

We imaged an area which includes both AGC~229384 and AGC~229385 on the
night of 6 February 
2013 with nine dithered 300 second exposures in each of the $g'r'i'$
filters. We observed AGC~229383 on 2 May 2014 with nine 
dithered 300 second exposures in both the $g'$ and $r'$ filters. By
combining data from these two nights of observations, we have
contiguous deep multi-wavelength imaging coverage over an area
$\sim$$40' \times 40'$. 
We also imaged this field with an $80$\AA \, narrow-band H$\alpha$
filter during photometric conditions on the night of 6 February
2013. Our dithered sequence of 
nine 300s images, while not calibrated, do not show any H$\alpha$
detections at the locations of the three HI sources in the HI1232+20
system, but do show a background spiral galaxy (AGC~222741) quite
clearly.

We reduced our observations using the QuickReduce (QR, \citealt{kotulla14})
data reduction pipeline, and supplemented this processing with an
additional illumination correction. QR was run interactively
in the One Degree Imager Pipeline, Portal, and Archive
(ODI-PPA)\footnote{The ODI Pipeline, Portal, and Archive (ODI-PPA) is
  a joint development project of the WIYN Consortium, Inc., in
  partnership with Indiana University's Pervasive Technology Institute
  (PTI) and with the National Optical Astronomy Observatory Science
  Data Management (NOAO SDM) Program.}
science gateway (\citealt{gopu14}, \citealt{young13}). The PPA interface
allows the user to select which observations will be reduced, and runs
all of the reductions on computing resources at the Pervasive
Technology Institute (PTI) at Indiana
University.

The QR pipeline includes: masking
of saturated pixels, crosstalk, and persistence; overscan
subtraction; bias level subtraction; dark current subtraction;
non-linearity corrections to each cell; flat field correction from
dome flat fields; cosmic-ray removal; fringe removal (in $i'$); pupil
ghost correction. However, the final pipeline-processed data still
have uncorrected instrumental artifacts in them, especially at very
faint intensity levels. In order to produce images that are suitable
for low surface brightness analysis, we need to correct for the small
gradients, sky level offsets, and other artifacts in particular
cells. Once these effects are corrected, the dithered images can be
combined into a final deep image.

In order to remove these image artifacts, we apply an illumination
correction using dark sky flats generated from the observations
themselves. For a particular filter, we mask all objects in the
images, then use a median algorithm to combine all of the exposures
into a dark sky flat field, which is then smoothed with a
$3$$\times$$3$-pixel smoothing element. Each exposure is then divided by
this illumination correction image. 

Before combining all exposures in a dither pattern, we re-project them
to a common pixel scale and also scale the images to a common flux
level using 
measurements of stars in the field and SDSS DR9 catalog magnitudes
\citep{ahn12}. This compensates for varying
sky transparency during the dither sequence, and typically yields
final photometric zeropoints with standard deviations of $0.02-0.03$
magnitudes. The $g'$ filter calibrations required a $g'$-$i'$ color term
of amplitude $0.079 \pm 0.013$, but $r'$ and $i'$
calibrations required no color term. The point sources in our final
combined images have an average FWHM of $0.7'' - 0.9''$.

We also create a deep ``detection-only'' image by combining all images
from both pointings in all filters to reach the faintest light
possible. This detection image is then binned to $1.2''$ resolution to
bring out very faint emission, and is shown in Figure~\ref{allim} with
relevant HI sources labeled and HI synthesis contours overlaid. The
contrast levels in this image have
been stretched to show the exquisite sensitivity to faint 
light. In this view, the optical counterpart to AGC~229385 is
strikingly visible, as will later be discussed. Also visible in the
upper left corner of the image is diffuse 
filamentary emission from Galactic cirrus. This foreground emission comes from
reflections of star light off cold dust clouds in our
Galaxy (\citealt{sandage76}, \citealt{witt08}). Multiple infrared surveys
(IRAS, \citealt{sfd98}, WISE, \citealt{wright10}) also observe this
dust via its thermal emission, and show features that are coincident
with the faint optical emission we see in our image. It even shows up
weakly in the overlapping deep archival GALEX UV image. Galactic cirrus
is very faint and diffuse at optical wavelengths and typically only
visible in deep, wide-field images that are very accurately
flat-fielded (e.g., \citealt{rudick10}).

\begin{figure*}[htb]
\centering
\includegraphics[width=17.5cm]{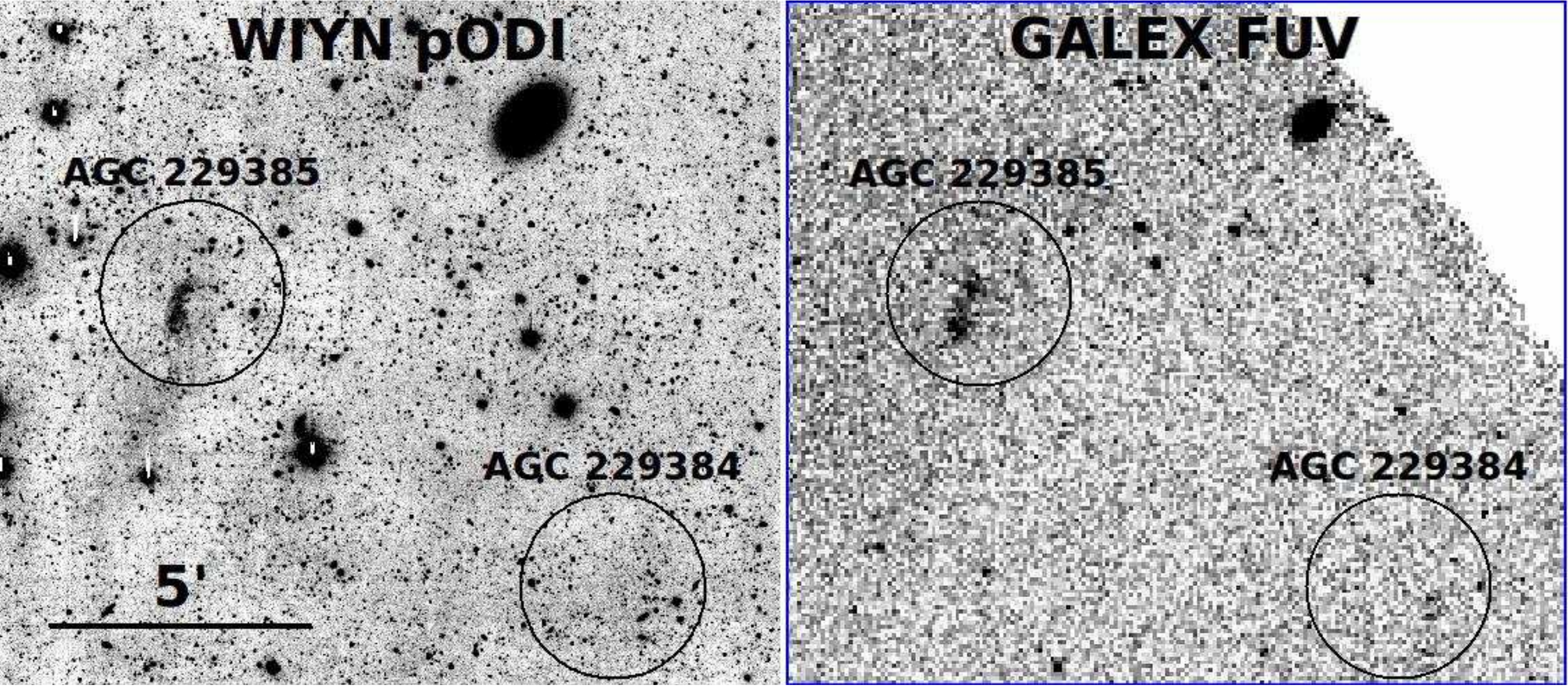}
\caption{Comparison of WIYN pODI image and GALEX FUV
  image. The pODI detection image is binned $12\times12$ so that
  pixels are $1.3''$ on a side. The FUV image has been binned
  $3\times3$ so that pixels are $13''$ on a side.
  Two of the three HI sources in the HI1232+20 system are labeled on
  both panels with $3'$ circles, the size of the ALFALFA
  beam. AGC~229385 clearly has an optical and ultraviolet 
  counterpart while AGC~229384 lacks a counterpart at either
  wavelength. AGC~229383 has not been observed by
  GALEX. 
\label{uv}
}
\end{figure*}

\subsection{HI synthesis imaging with WSRT}

\label{obs.hi}

We observed the HI1232+20 system with four 12h pointings at the
Westerbork Synthesis Radio Telescope (WSRT), three of which were
centered on (12:31:52.6~+20:22:59) to encompass the centroids of
AGC~229385 and AGC229384, and one of which was centered on
(12:31:08.5~+20:31:41.9), to encompass AGC~229383. The
primary beams of the two pointing centers are $35'$ wide and cover
nearly all of the area displayed
in Figure~\ref{allim}. We observed the HI line in one band with 10~MHz
bandwidth, two polarization products, and 1024 channels, ensuring a
broad range of line free channels for continuum subtraction and a
velocity resolution of $4.12$~\kms \, after Hanning smoothing.

The data were reduced using the same automated data reduction pipeline
as applied in \citet{wang13}, originally used by \citet{serra12},
using the data reduction software Miriad \citep{sault95} wrapped into
a Python script. The data were automatically 
flagged for radio interference using a clipping method after filtering
the data in both the frequency- and time-domain. After the primary
bandpass calibration, the data were iteratively deconvolved with the
CLEAN algorithm, using clean masks determined on the cube with
decreasing clip levels, to then apply a self-calibration. The
calibration solution was applied to the visibilities and the continuum
was subtracted in the visibility domain to then invert the data after
Hanning smoothing, using a set of combinations of Robust weighting and
tapering with a Gaussian kernel, as well as a binning in the frequency
domain. Finally the data cubes were iteratively cleaned using clean
masks determined by filtering the data cubes with Gaussian kernels and
applying a clip level. The clean cutoff level was set to the rms noise
in the data cubes. Because we hence cleaned the data comparably
deeply, no correction of the intensity levels of the residuals was
made. 

Our pipeline produces
cubes at each centroid with three different robustness weightings,
$r=0.0$, $r=0.4$, and $r=6.0$, binned to a velocity resolution of
$6.2$ \kms\ after Hanning smoothing ($12.4$ \kms \, for the $r=6.0$
cube). The noise level in the cubes for the three respective
robustness weightings are $0.40$, $0.36$, and $0.24$ mJy/beam/channel,
with beam sizes of $39''$$\times$$13''$, $45''$$\times$$15''$, and
$54''$$\times$$20''$.

For each cube we then created HI total flux maps by summing masked
cubes along the velocity axis. We created the masks by smoothing the
images to twice the beam size, and then keeping any pixel
3$\sigma$ above the noise level. From these we calculate HI column
density maps assuming optically thin HI gas such that
N$_{HI}$$=$$1.823\times10^{18} \int T_b dv$ cm$^{-2}$. Since the final
contour map results from the combination of multiple WSRT
observations ($3$ at the SE pointing, and $1$ at the NW pointing), the
signal-to-noise ratio varies across the image and is less sensitive
near AGC~229383. The lowest HI contour shown on the images,
$1$$\times$$10^{19}$~cm$^{-2}$, corresponds to a less significant
detection in the region around AGC~229383 than it does in the region
around AGC~229385 and AGC~229384. As a result, there were locations
outside of the main signal from AGC~229383 where the HI column density
exceeded $1$$\times$$10^{19}$~cm$^{-2}$, but since the significance of
the detection was $<$$3\sigma$, those contours are not shown.

We additionally create a one dimensional {integrated} HI line
profile for each 
object, as displayed in Section \ref{results}. 
We fitted the line with both the two horned function applied 
in the ALFALFA data processing, and using a standard Gaussian fit and
note that the fluxes from the fits match well within random errors. 
We recover $99$\% of the ALFALFA flux in the WSRT spectrum of
AGC~229385, and $92$\% in AGC~229384, but only $52$\% in
AGC~229383. The spectra for AGC~229385 and AGC~229384 are both well
fitted by a Gaussian profile, and though both may show slight
deviation from Gaussian, using two Gaussians does not return a better
result. The spectrum for AGC~229383 is not well fitted by either a
Gaussian or a two-horned fit.

\begin{deluxetable*}{lccccc}
\tablewidth{0pt}
\tablecaption{Derived properties and limits from observations \label{abs}}
\tablehead{
\colhead{Quantity [units]} & \colhead{AGC~229383} &
\colhead{AGC~229384} & \colhead{AGC~229385}  }
\startdata
$m_{g'}$  [mag] & $>20.7$ & $>20.7$ & $19.20$ ($0.03$) \\
$m_{r'}$  [mag] & $>20.2$ & $>20.5$ & $19.27$ ($0.03$) \\
$m_{i'}$  [mag] & \nodata & $>19.7$ & $19.36$ ($0.04$) \\
$\mu_{g',{\rm peak}}$  [\magsec] & $>27.8$ & $>27.9$ & $26.4$ ($0.1$) \\
$\mu_{r',{\rm peak}}$  [\magsec] & $>27.3$ & $>27.7$ & $26.5$ ($0.1$) \\
$\mu_{i',{\rm peak}}$  [\magsec] & \nodata & $>26.8$ & $26.1$ ($0.1$) \\
\hline
Optical major axes (at $\mu_{g'}$$=$$27$) [kpc] & \nodata & \nodata &  $7 \times 3$  \\
HI major axes (at $5$$\times$$10^{19} $cm$^2$) [kpc] & $12 \times 4$ & $14 \times 10$ & $28 \times 16$ \\
\hline
$M_{g'}$           [mag] & $>-11.3$ & $>-11.3$ &  $-12.89$  \\ 
$M_{r'}$           [mag] & $>-11.8$ & $>-11.5$ &  $-12.79$ \\ 
$M_{i'}$           [mag] & \nodata  & $>-12.3$ &  $-12.69$ \\ 
$g'-r'$           [mag] &  \nodata & \nodata & $-0.09$ \\ 
$B-V$            [mag] & \nodata & \nodata & $0.13$ \\
$M_B$            [mag] & $>-11.2$ & $>-11.1$ & $-12.72$ \\
\hline
$L_{FUV}$        [$L_\odot$] & \nodata & \nodata & $2.41  \times 10^7$ \\
$L_{NUV}$        [$L_\odot$] & \nodata & \nodata & $2.66  \times 10^7$ \\
$SFR_{NUV}$      [M$_\odot$/year] & \nodata & \nodata & $4.1  \times 10^{-3}$ \\
$SFR_{FUV}$      [M$_\odot$/year] & \nodata & \nodata & $6.9  \times 10^{-3}$ \\
\hline
log $M_{HI}$     [log $M_\odot$] & $8.08$  & $8.30$  & $8.86$  \\
$M_\star$        [M$_\odot$] & $<$$3.7 \times 10^5$  &  $<$$3.4 \times 10^5$  &  $1.5 \times 10^6$ \\
$M_{HI}/M_\star$    & $>320$ & $>580$ & $290$ \\
$M_{HI}/L_{g'}$     [$M_\odot / L_\odot$] & $>31$ & $>57$  & $45.8$ \\
$M_{HI}/L_B$       [$M_\odot / L_\odot$] & $>26$ & $>48$  & $38.2$  
\enddata
\tablecomments{Apparent magnitudes ($m_{g'}$, $m_{r'}$, $m_{i'}$) are
  not corrected for Galactic extinction. 
  Absolute magnitudes
  ($M_{g'}$, $M_{r'}$, $M_{i'}$, $M_B$) luminosities ($L_{FUV}$,
  $L_{NUV}$), and colors ($g'-r'$, $B-V$) are corrected for Galactic
  extinction from \citet{sf11}. All absolute quantities assume a
  distance of $25$~Mpc. $M_B$ and $B-V$ are determined from
  conversions in \citet{jester05}. 
  Upper limits are determined where sources are not detected in pODI
  observations and are at $3\sigma$ confidence levels. Uncertainties
  on measured quantities are indicated in parentheses.}
\end{deluxetable*}

Finally, we produce HI velocity maps using two different methods. We
created standard moment 1 maps from cubes masked at 3$\sigma$, and
additionally fitted Gaussian functions to each individual profile in
the datacube using the GIPSY task XGAUFIT. The resulting maps from the
two methods are virtually identical, and we show the maps in 
Section~\ref{results}.

To further analyze the velocity field we create Position-Velocity (PV)
diagrams for each source. We produced the PV diagrams by taking an
$18''$ ($1$ minor axis beam width) wide slice along the HI major axis 
centered on the HI surface density centroid. We measured the position
angle and centroid from the surface density profile since the
variations in the velocity field leave the major velocity axis and
center uncertain. We note that none of the PV diagrams change
significantly for small variations in position angle or slice width.


\subsection{Archival GALEX observations}
\label{galex}

AGC~229385 (the strongest HI detection of the HI1232+20 system) had a
very faint 
UV counterpart visible in an archival dataset from GALEX
(\citealt{martin05}, \citealt{morrissey07}). GALEX obtained images in
the far ultra-violet 
(FUV) from $1344$ to $1786$\AA \, with $4.3''$ FWHM resolution, and in
the near ultra-violet (NUV) from $1771$ to $2831$\AA \, with $5.3''$
resolution. These UV images are especially sensitive to young
stellar populations, and should help to identify sites of recent star
formation.

AGC~229385 was imaged by GALEX in May 2007 in the NUV and FUV
bands, with exposure times of 1145s in both bands. 
A set of matched images is shown in Figure~\ref{uv}, with data from
pODI WIYN and GALEX FUV.
The locations of AGC~229385 and AGC~229384 are shown in these images,
while AGC~229383 lies outside 
of any archival GALEX image. The brightest HI source, AGC~229385, is
faintly visible in the UV image as a diffuse source. However, the
GALEX pipeline (GR7, Bianchi \etal 2014) does not identify this
diffuse object as a source, instead shredding it into multiple point
sources. We measure the brightness of AGC~229385 in the FUV and NUV
images ourselves in an aperture matched to our optical images, as
discussed in Section~\ref{results}. There is no source visible in the
FUV image at the position of AGC~229384.

\section{Results} \label{results}

In the following sub-sections we describe the
results of our followup observations for the HI1232+20 system. The
derived results are summarized in Table~\ref{abs}. We also consider the environment around this
system, and uncertainties in the adopted distance.

\subsection{AGC~229385}
\label{AGC229385}

\begin{figure*}[!htb]
\centering
{\bf a \hspace{2.2in}  b \hspace{2.2in} c }\\
\includegraphics[width=5.8cm]{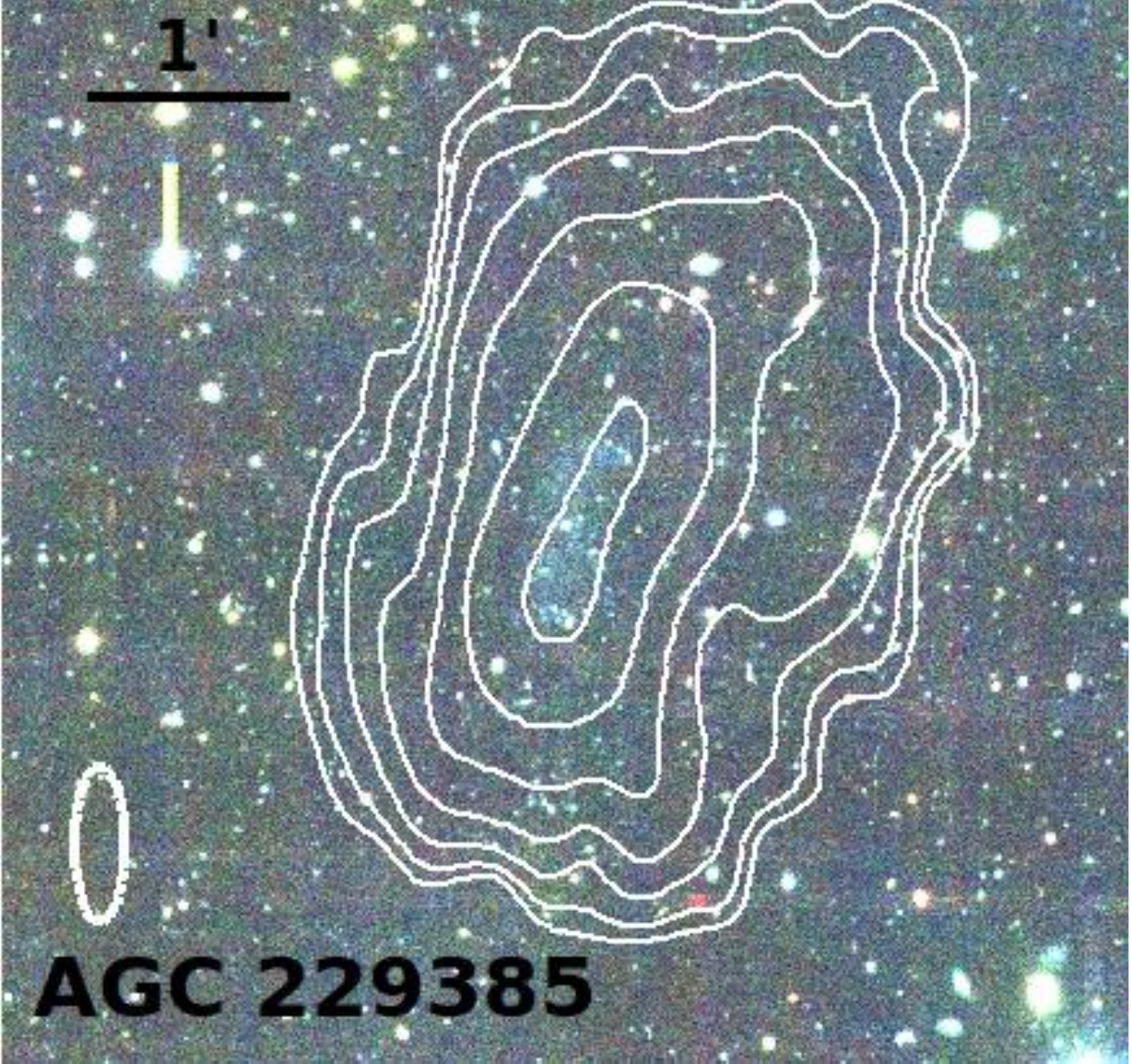}
\includegraphics[width=5.8cm]{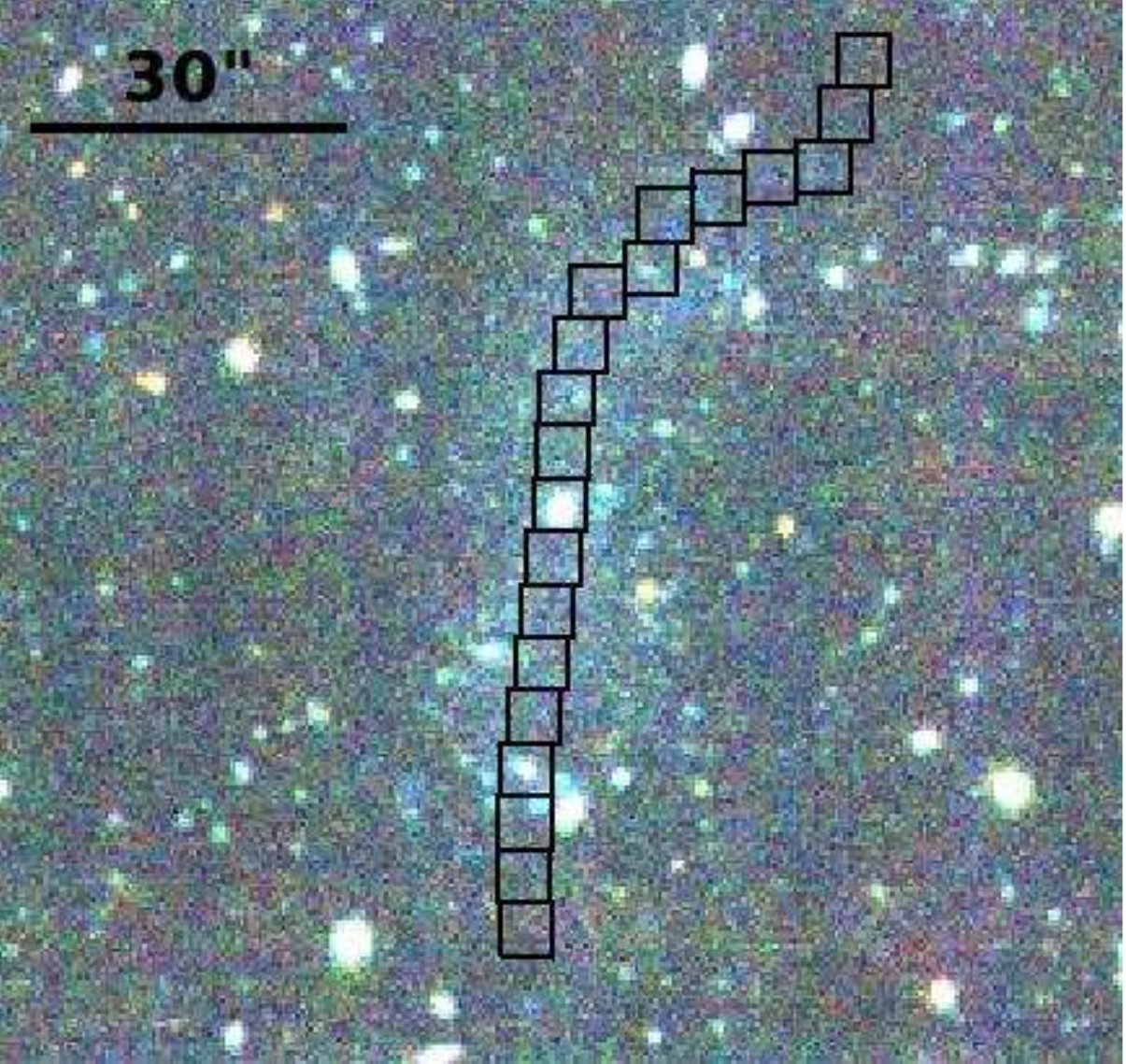} 
\includegraphics[width=5.8cm]{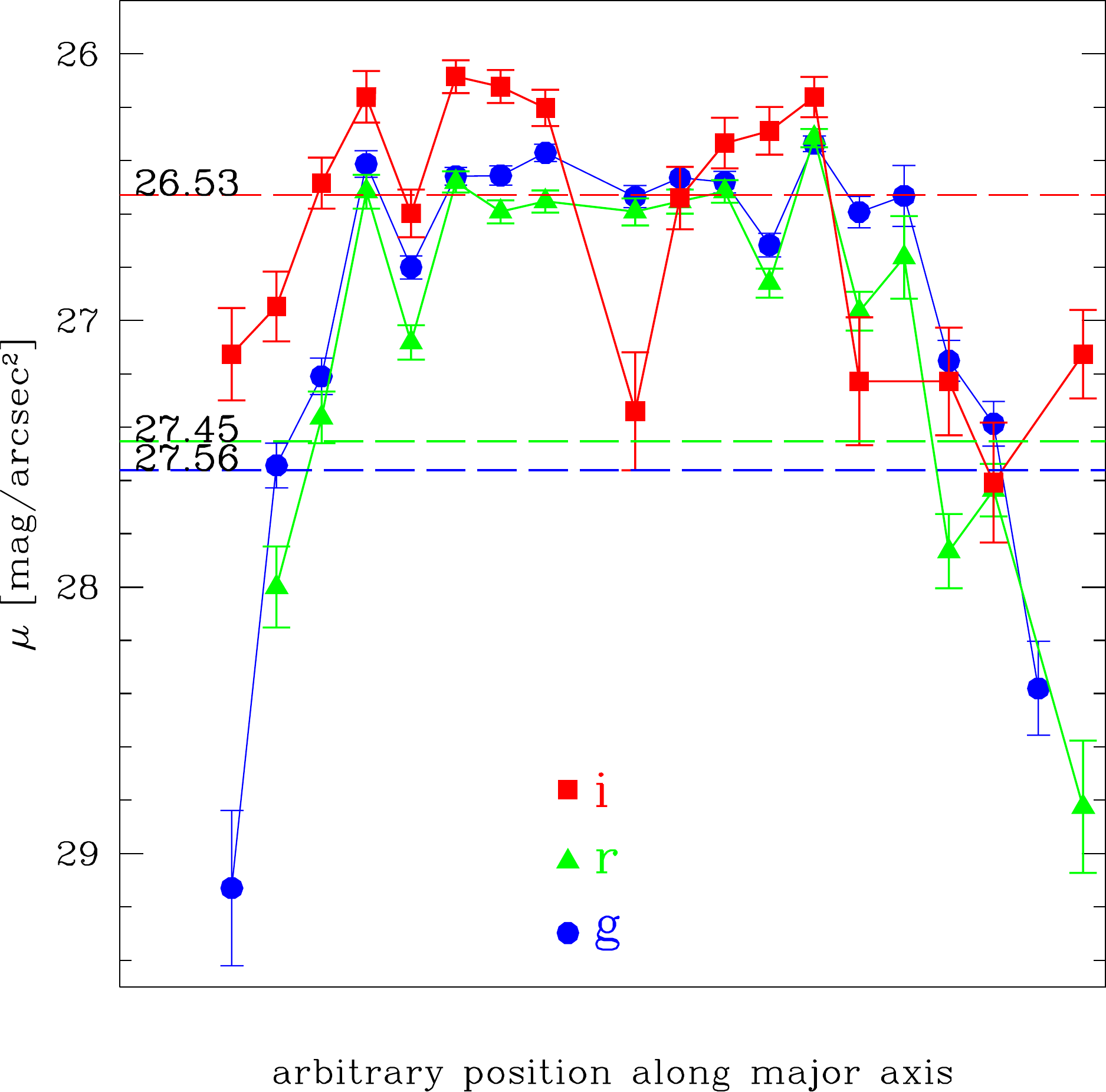} \\
{\bf d \hspace{2.2in}  e \hspace{2.2in} f }\\
\includegraphics[width=5.6cm]{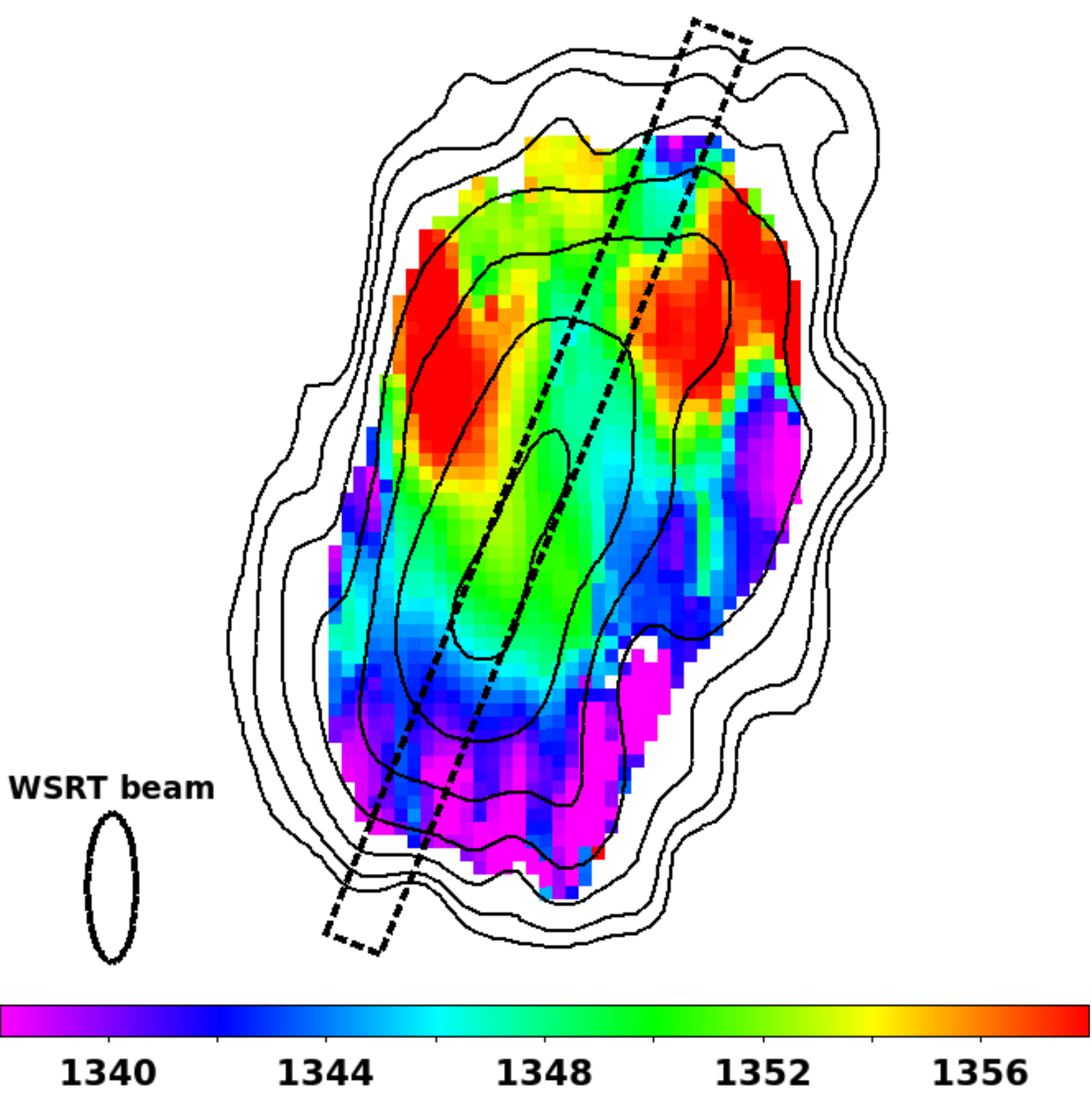}
\includegraphics[width=5.6cm]{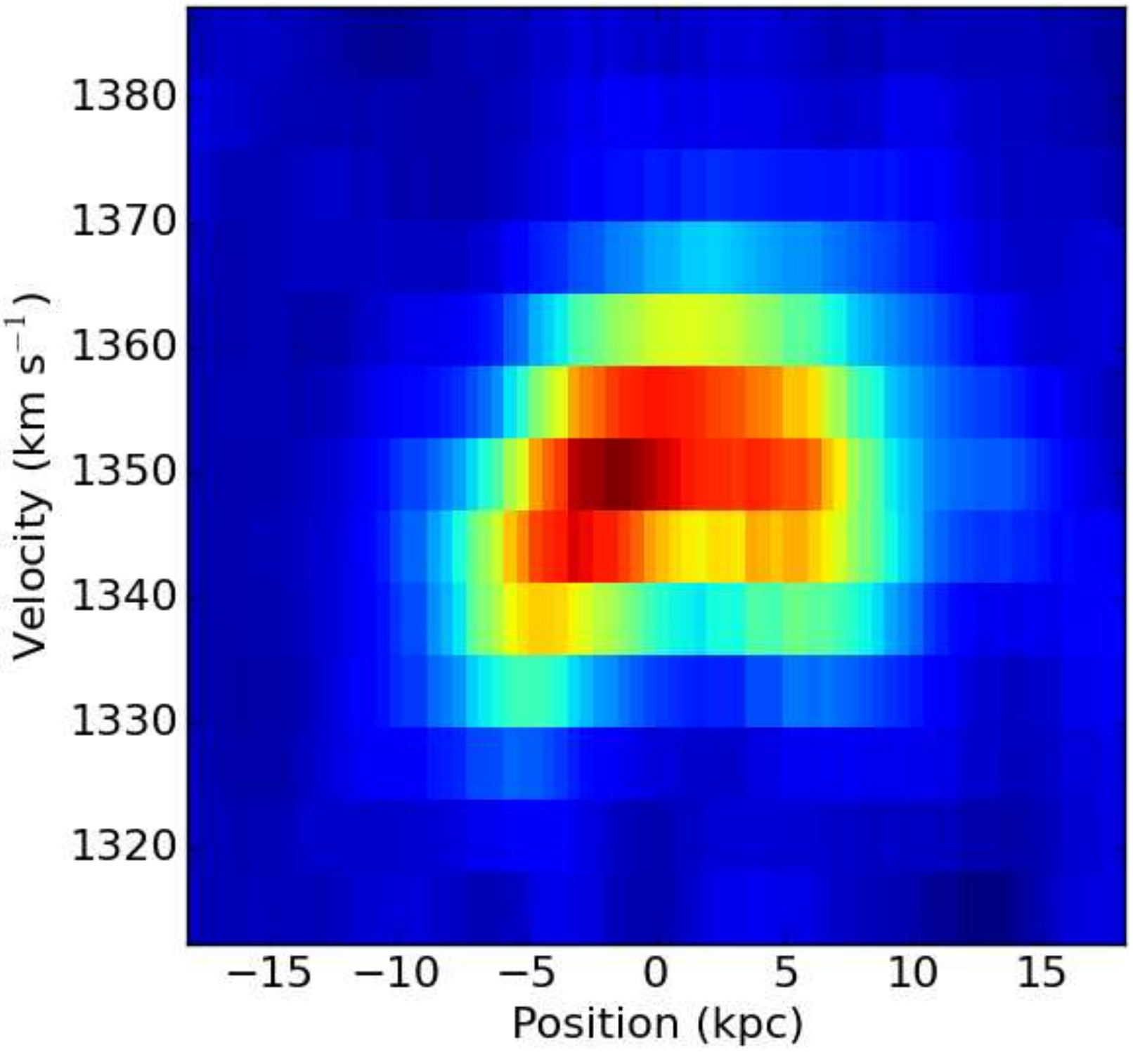}
\includegraphics[width=6.6cm]{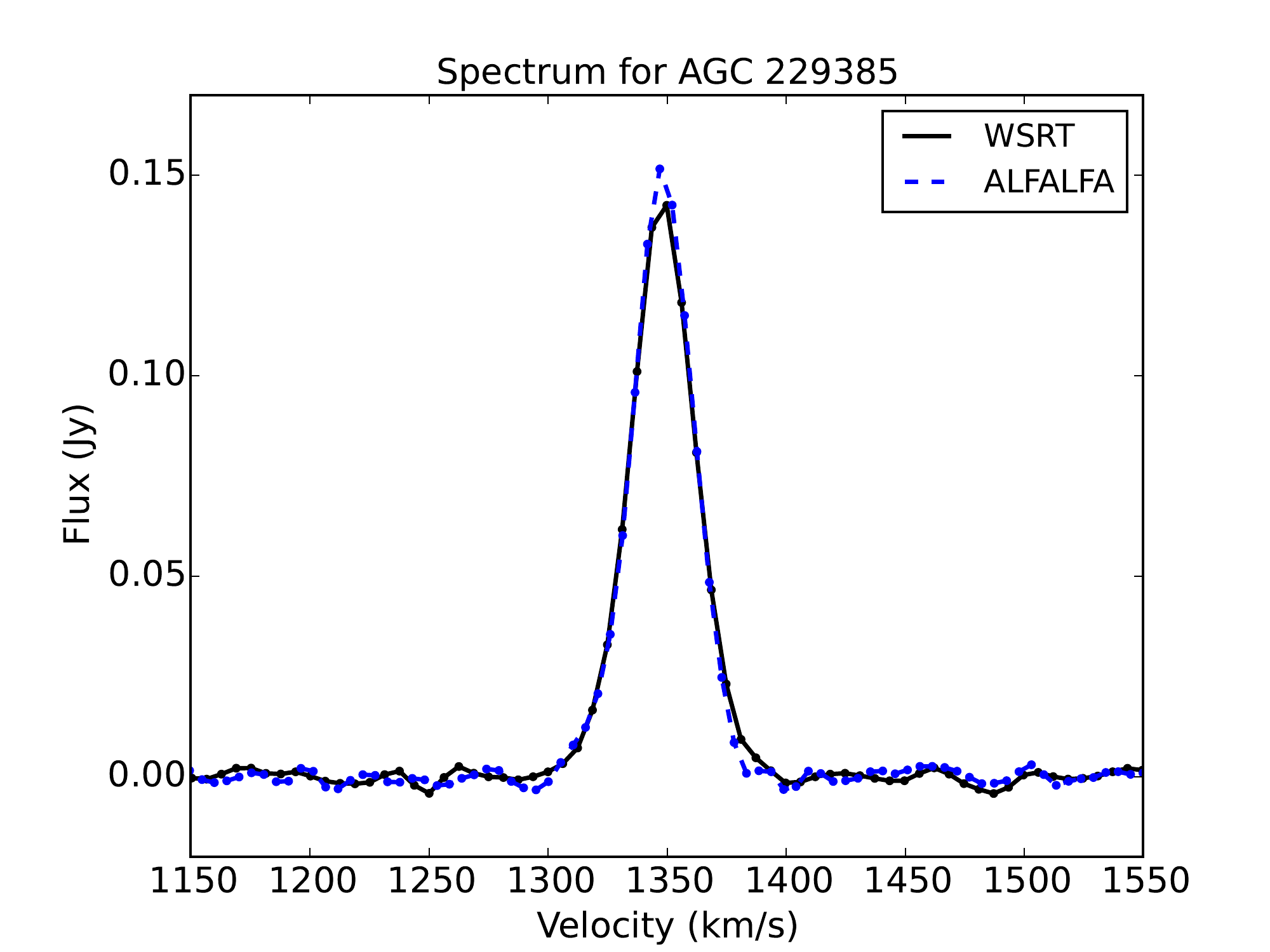} \\
\caption{
  Our observations of AGC~229385.
{a)}~3-color optical image from WIYN pODI binned $4 \times 4$
  so that pixels are $0.44''$ on a side. North is up, East is left, and
  the angular scale is indicated on the image. White contours of HI
  emission are 
  from WSRT and are at are at $1$, $2$, $4$, $8$, $16$, $32$, and $64
  \times 10^{19}$   atoms cm$^{-2}$, and the white ellipse at the
  bottom left shows the size and shape of the WSRT beam.
{b)}~Zoomed-in view of the WIYN pODI observations, with black
  boxes indicating regions used for surface brightness measurements. 
{c)}~Surface photometry for AGC~229385 from pODI images in
   $g'$, $r'$, and $i'$ filters, measured in $5'' \times 5''$ boxes
   following the  brightest part of the optical emission, after
   masking obvious stars and background galaxies. Dashed lines show
   the $3\sigma$ surface brightness detection limits in each
   filter. Blue points and lines indicate $g'$ filter measurements,
   green indicate $r'$ measurements, and red indicate $i'$
   measurements.
{d)}~XGAUFIT velocity field from WSRT with the same HI contours
  as above. The ellipse in the bottom left corner indicates the size
  and shape of the WSRT beam. The dashed rectangular region shows the
  slice along the major axis which is used to generate the P-V
  diagram.
{e)}~Position-Velocity diagram from WSRT. 
{f)}~ALFALFA and WSRT HI spectra of AGC~229385. As discussed in
  the text, HI fluxes from ALFALFA and WSRT are in good agreement. Blue
  dashed lines represent the ALFALFA spectra, while the solid black
  lines show the WSRT $r=0.4$ spectra.
\label{385}
}
\end{figure*}
Figure~\ref{385} shows our observations of AGC~229385.
Figure~\ref{385}a
is a deep 3-color image from WIYN pODI with contours from our WSRT HI
synthesis map at the highest resolution ($39''$$\times$$13''$). Figure
\ref{385}b shows a zoomed-in region around the optical 
counterpart with the regions we use to measure its surface
brightness indicated by black squares. The 
optical emission from AGC~229385 appears very blue, and coincides 
spatially with the peak of the HI distribution. The optical component
appears $\sim$$5\times$ less extended than the radio emission, but
both are similarly elongated in the northeast-southwest direction.

AGC~229385 has an unusual optical morphology that is not simple to 
describe. The optical counterpart is elongated in the N-S direction,
but has a nearly constant surface brightness across its entire
extent. We fitted an ellipse to the $5\sigma$ contour on the $g'$ image 
and found a semi-major axis of $32''$ and semi-minor axis of
$10''$. This $5\sigma$ ellipse has a position angle of $15^\circ$
(measured clockwise from N) and an ellipticity ($\epsilon$$=$$1 - b/a$)
of $\epsilon$$=$$0.68$. 

After attempts to fit elliptical annuli to the optical images
of AGC~229385
resulted in inconsistent and divergent surface brightness profiles, we
decided to measure the surface brightness in small regions instead. We
measured the optical surface brightness of AGC~229385 in
$5'' \times 5''$ regions following the curving shape of 
this source from south to north, after masking all obvious foreground
and background sources. These regions are shown as black boxes in
Figure~\ref{385}b. We also placed similar regions 
outside the periphery of the source to determine the local sky
value. The surface brightness traces along AGC~229385 are
plotted in the Figure~\ref{385}c.
We calculated the surface brightness level that corresponds to $3$
times the standard deviation in the sky, and label it as the $3\sigma$
detection threshold. The $g'$ surface brightness profile has the
highest signal-to-noise ratio, and is well above the $3\sigma$
level. The $r'$ profile is weaker but still well-measured. In $i'$,
this source is only weakly detected, but still is above the $3\sigma$
surface brightness level. In all three filters a similar profile shape
is seen as the outlying regions show very little signal and the inner
regions show a relatively flat brightness distribution across the
source. To estimate a peak surface brightness value in each filter we
use the measurements from the three boxes just south of the bright
foreground star near the center of AGC~229385, where there are
relatively few contaminating sources which had to be masked and the
profiles are relatively smooth. The peak values
are calculated by averaging the measurements in these three boxes, and
are $26.4$, $26.5$, and $26.1$, in $g'$,$r'$, and $i'$,
respectively. While the formal uncertainties on these surface
brightness measurements are low ($\sim$$3-5\%$, owing to 
our accurate photometric calibrations and the good S/N of the optical
counterpart), the variations between adjacent boxes can be as high as
$0.1$~mag. Accordingly, we assign an uncertainty of $0.1$~mag to these
peak values.
We also measure photometry of this source in a $32''$ radius
aperture 
after masking obvious foreground and background sources. The results of the
surface and aperture photometry of all three sources
are summarized in Table~\ref{abs}.

\begin{figure*}[htb]
\centering
{\bf a \hspace{2.2in}  b \hspace{2.2in}  \, }\\
\includegraphics[width=5.6cm]{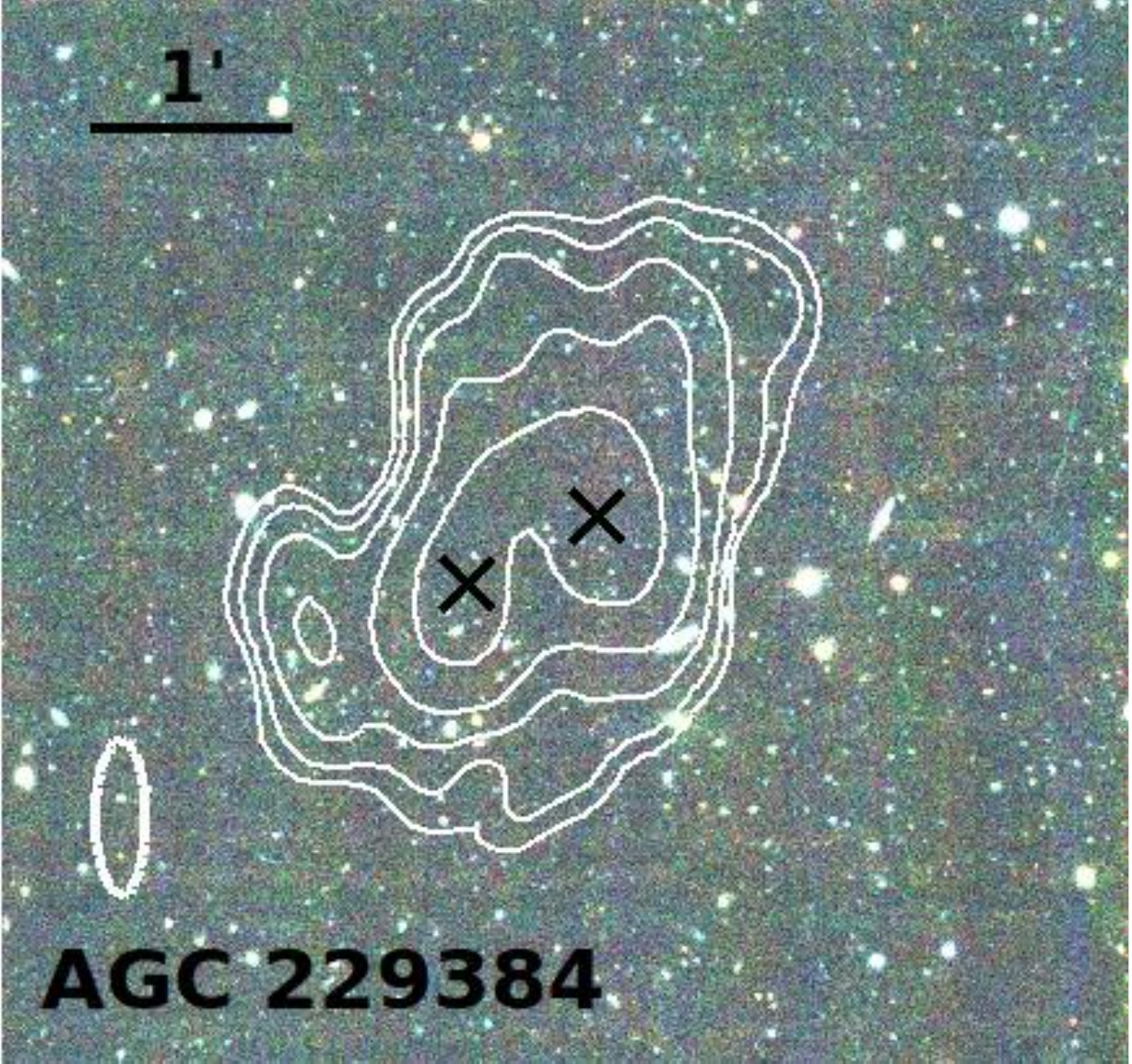}
\includegraphics[width=5.6cm]{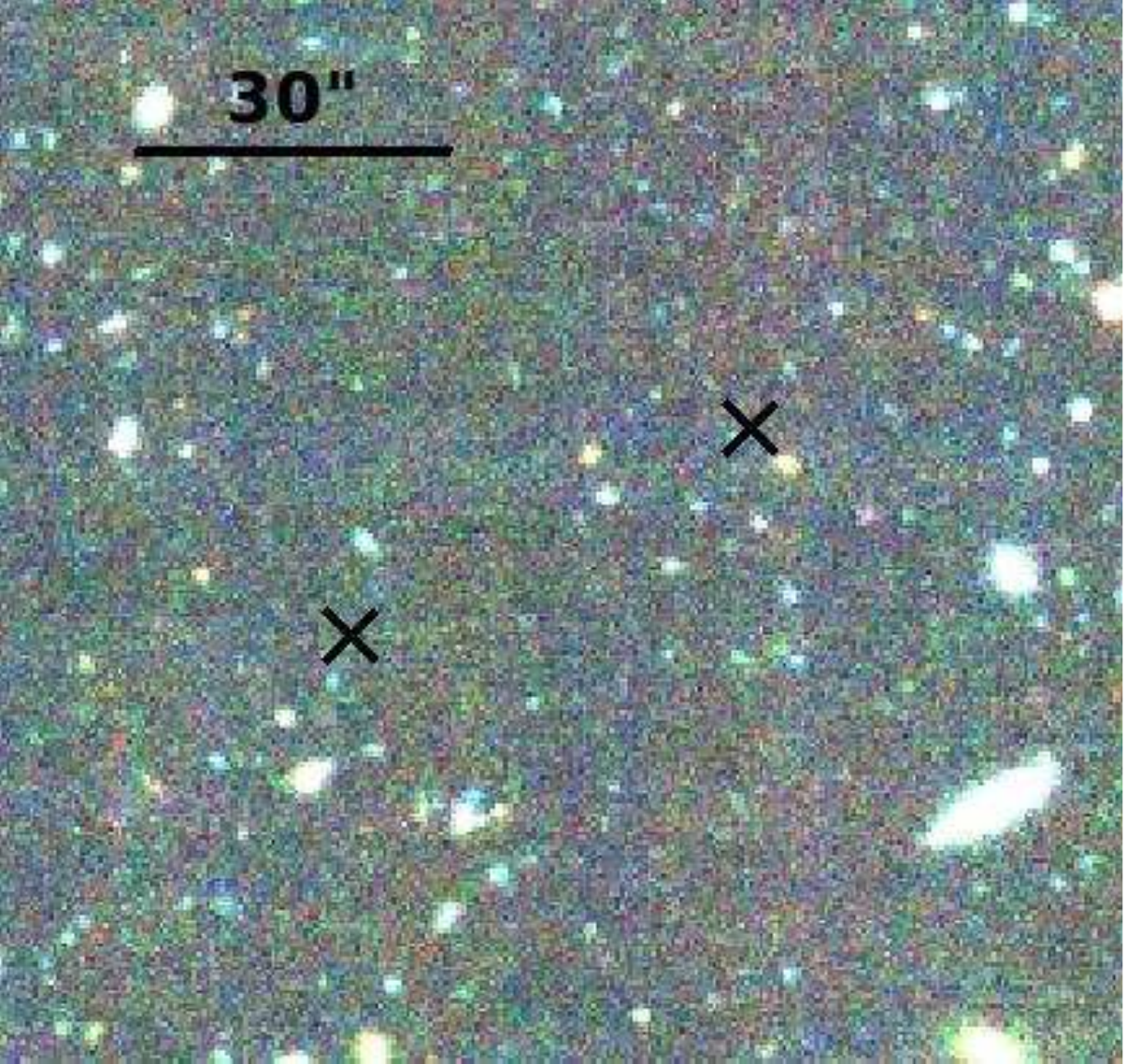} 
\hspace{5.6cm} .\\
{\bf d \hspace{2.2in}  e \hspace{2.2in} f }\\
\includegraphics[width=5.6cm]{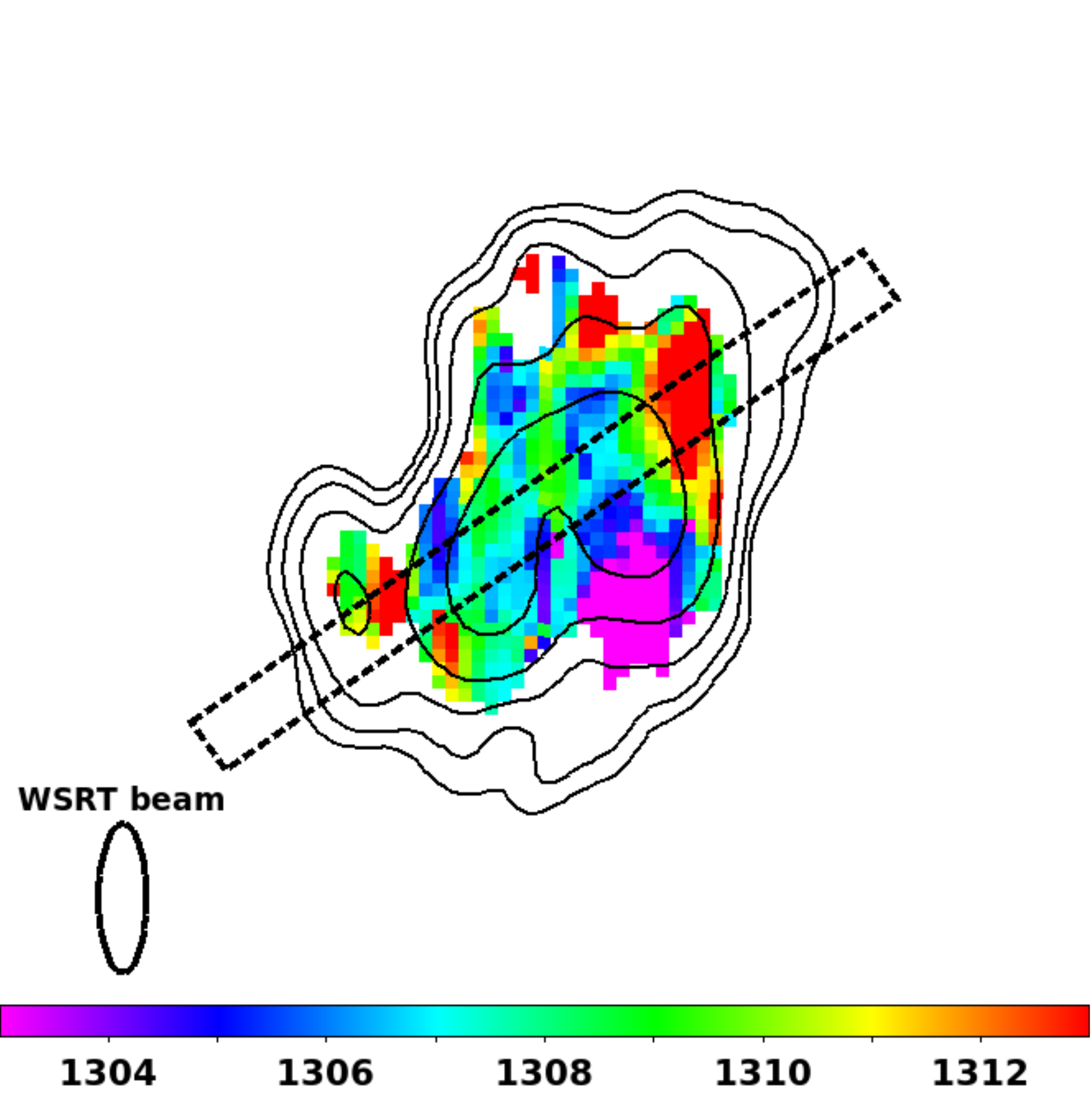}
\includegraphics[width=5.6cm]{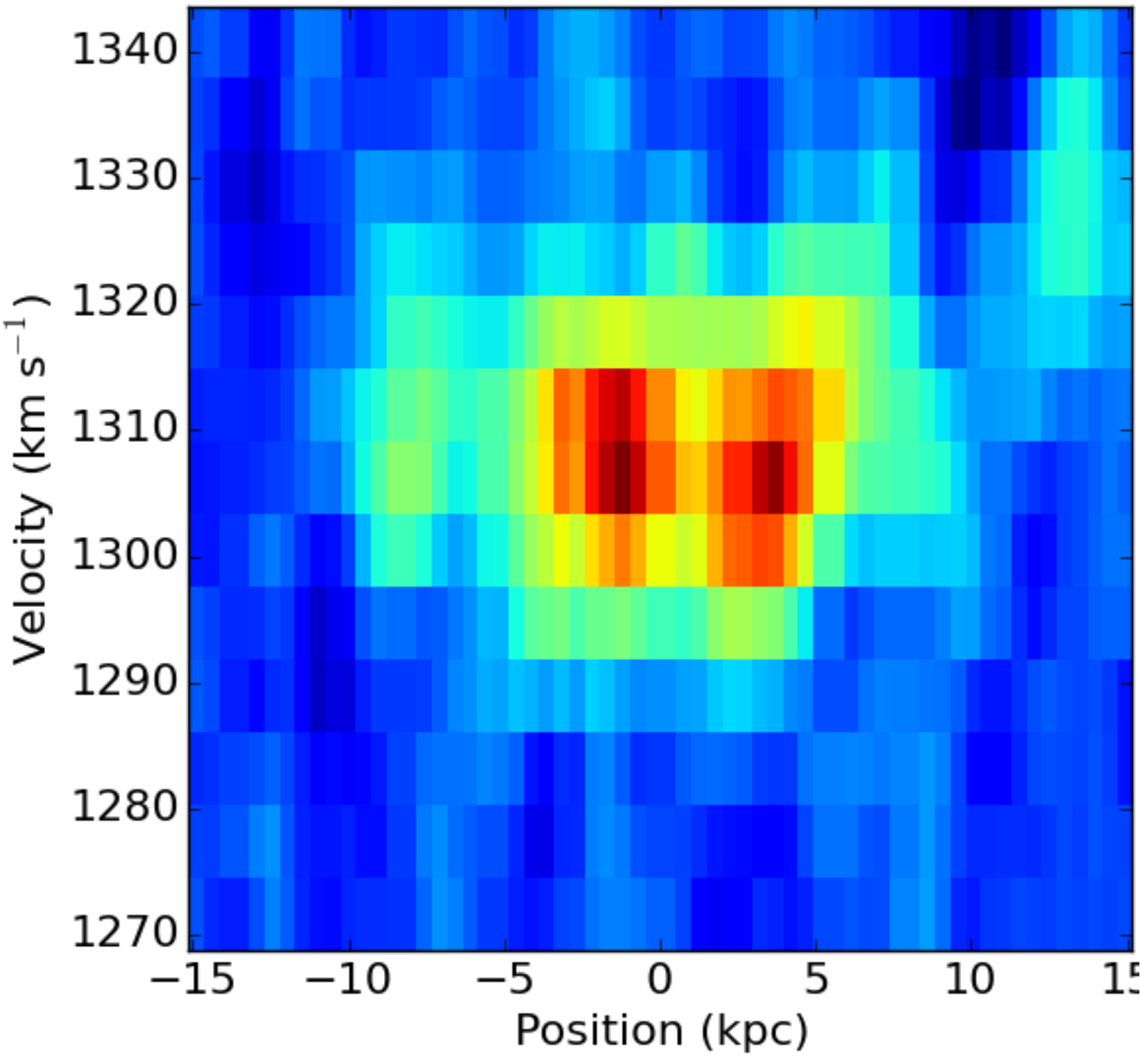}
\includegraphics[width=6.6cm]{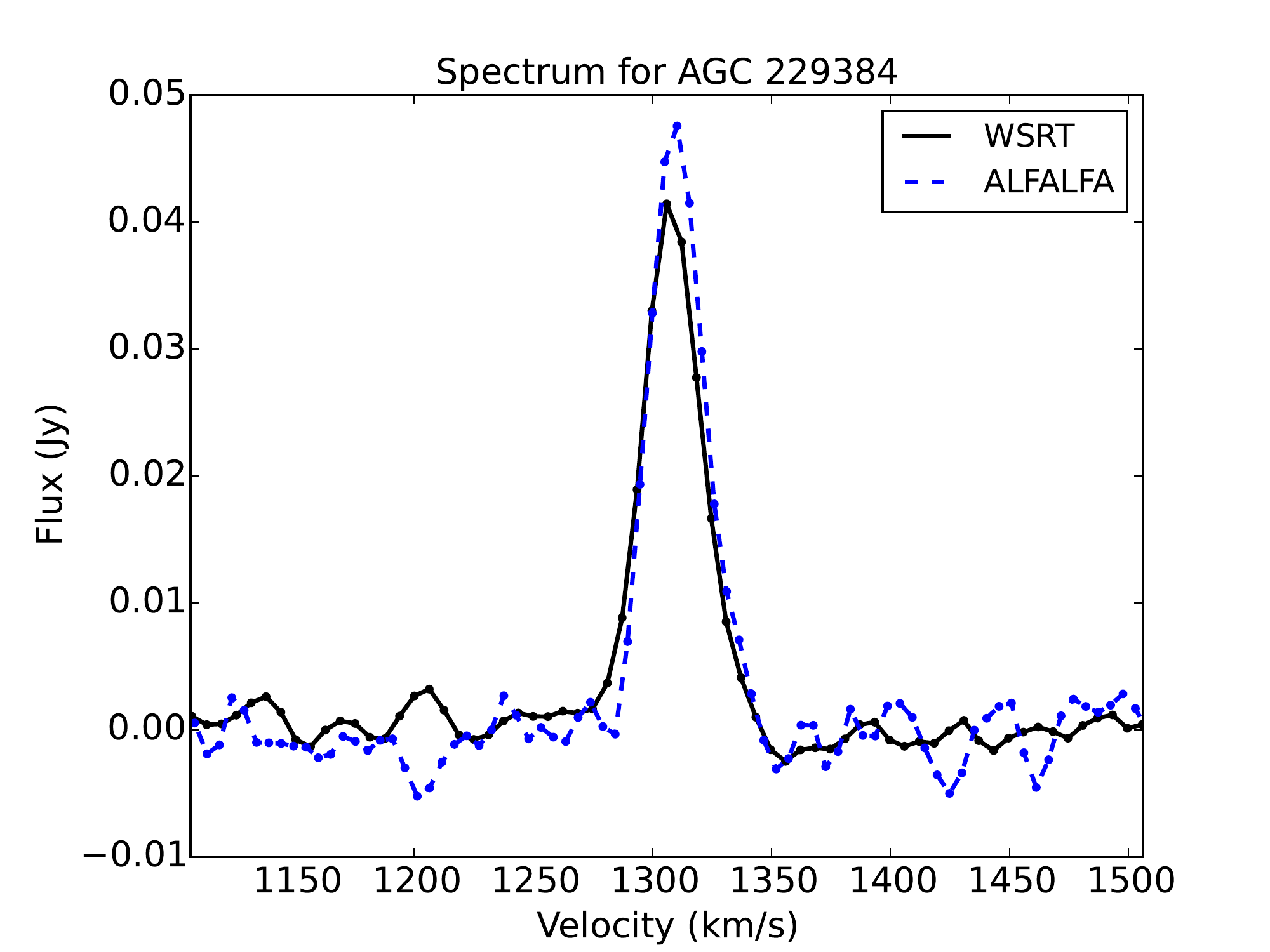}
\vspace{0.1in}
\caption{
  Our observations of AGC~229384, same panels as Figure
  \ref{385}. No optical counterpart is detected down to an upper limit
  of $\mu_{g'} \sim$$27.8$~\magsec, so no surface brightness profiles
  are shown. The black ``$\times$''s in panels~{a} and
  {b} mark the two 
  peaks of 
  the HI distribution. The HI velocity field of AGC~229384 in
  panel~{e} shows 
  no evidence of ordered rotation. As with AGC~229385, the WSRT and HI
  spectra in panel~{f} are in good agreement with each other.
\label{384}
}
\end{figure*}

The HI synthesis observations from WSRT (shown in the bottom row of
Figure~\ref{385}) are also difficult to interpret. The moment 0 map,
Figure~\ref{385}d, shows an HI source 
significantly more extended than its optical counterpart. 
AGC~229385 has HI  major and minor axes of $3.2'$$\times$$1.4'$
measured at an HI surface density of $1$~$M_\odot/$pc$^2$
(corresponding to a column density {of $12.5\times10^{19}$cm$^{-2}$}),
or $24$~kpc $\times 10$~kpc  
assuming a distance of 25~Mpc. WSRT measures significant emission
at lower column densities, out to $3.7' \times 2.2'$ or $28$~kpc
$\times 16$~kpc at $5 \times 10^{19} $cm$^{-2}$ and a furthest extent
of $\sim$$5'$ at $10^{19} $cm$^{-2}$.

The exact surface density profile of AGC~229385 is subject to 
its 3D geometry and inclination, which are difficult to determine
conclusively given our beam size and the unusual nature of
AGC~229385. AGC~229385 appears to have an HI position angle of
$21^\circ$ which gives $\sim$$6 \times 7$ resolution elements at the
furthest extent 
along the major and minor axes. We formally measure an inclination of
$63 \pm 4$ degrees assuming a thin HI disk and uncertainties of half
the beam size along the major and minor axes.

As an instructive exercise, we assume a disk geometry and compute
de-projected surface density profiles for AGC~229385 using
Robertson-Lucy deconvolution (\citealt{lucy74}, \citealt{warmels88},
the GIPSY task 
RADIAL). This method, developed for use in low resolution imaging,
works by collapsing the measured intensity along the minor axis, and
produces a one-dimensional profile, which is then iteratively matched
by a model one-dimensional profile produced by summing axisymmetric,
uniform density co-planar rings along lines of sight. This method does
not require knowledge of the inclination of the object, but still
assumes a disk geometry. The summed one-dimensional surface density
profile shows some asymmetry and two peaks with a slight depression in
the center. These features are reflected by asymmetry in the resulting
RADIAL model, and a strong suggestion of a hole in the center of the
HI distribution, a feature that would be smeared out in 2D ellipse
fitting analysis. If confirmed, this hole could be indicative of the
formation of cold atomic and molecular hydrogen in the center of the
object, or of a non-disky, more complicated HI distribution, possibly
caused by two recently merged components. However, higher resolution
data are necessary to confirm the existence of a gap in the HI
distribution.


\begin{figure*}
\centering
{\bf a \hspace{2.2in}  b \hspace{2.2in} c }\\
\includegraphics[width=5.6cm]{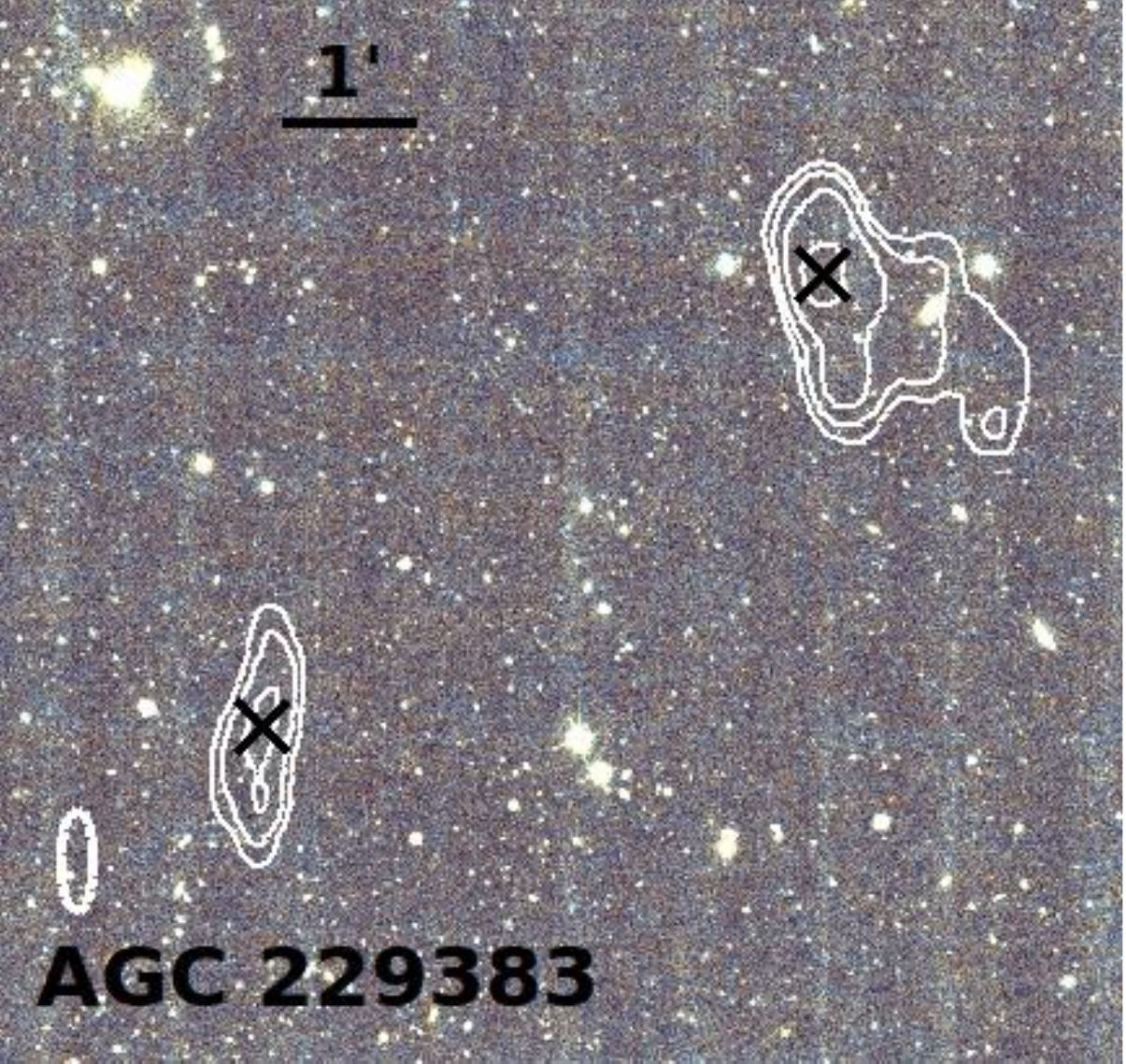}
\includegraphics[width=5.6cm]{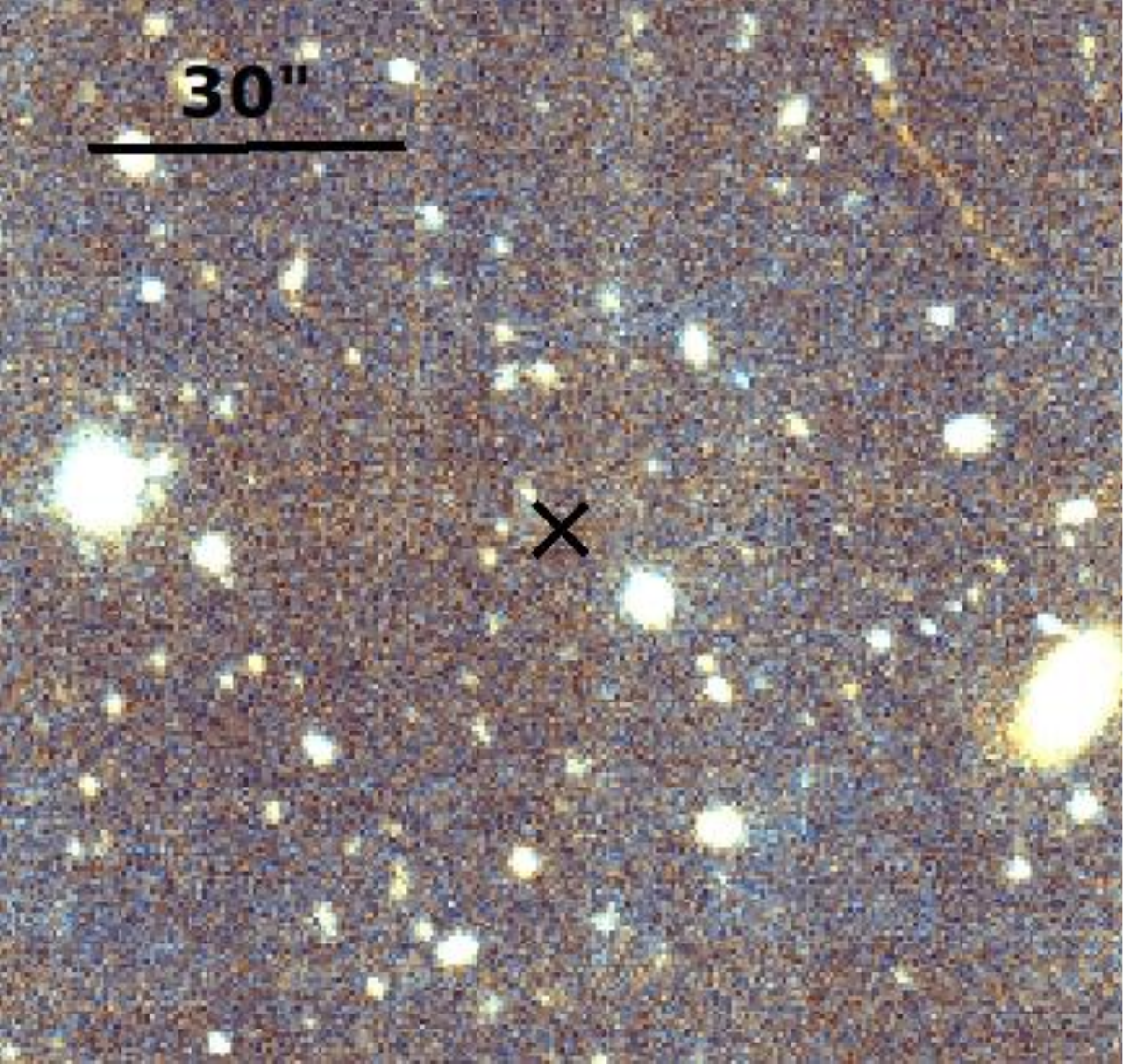}
\includegraphics[width=5.6cm]{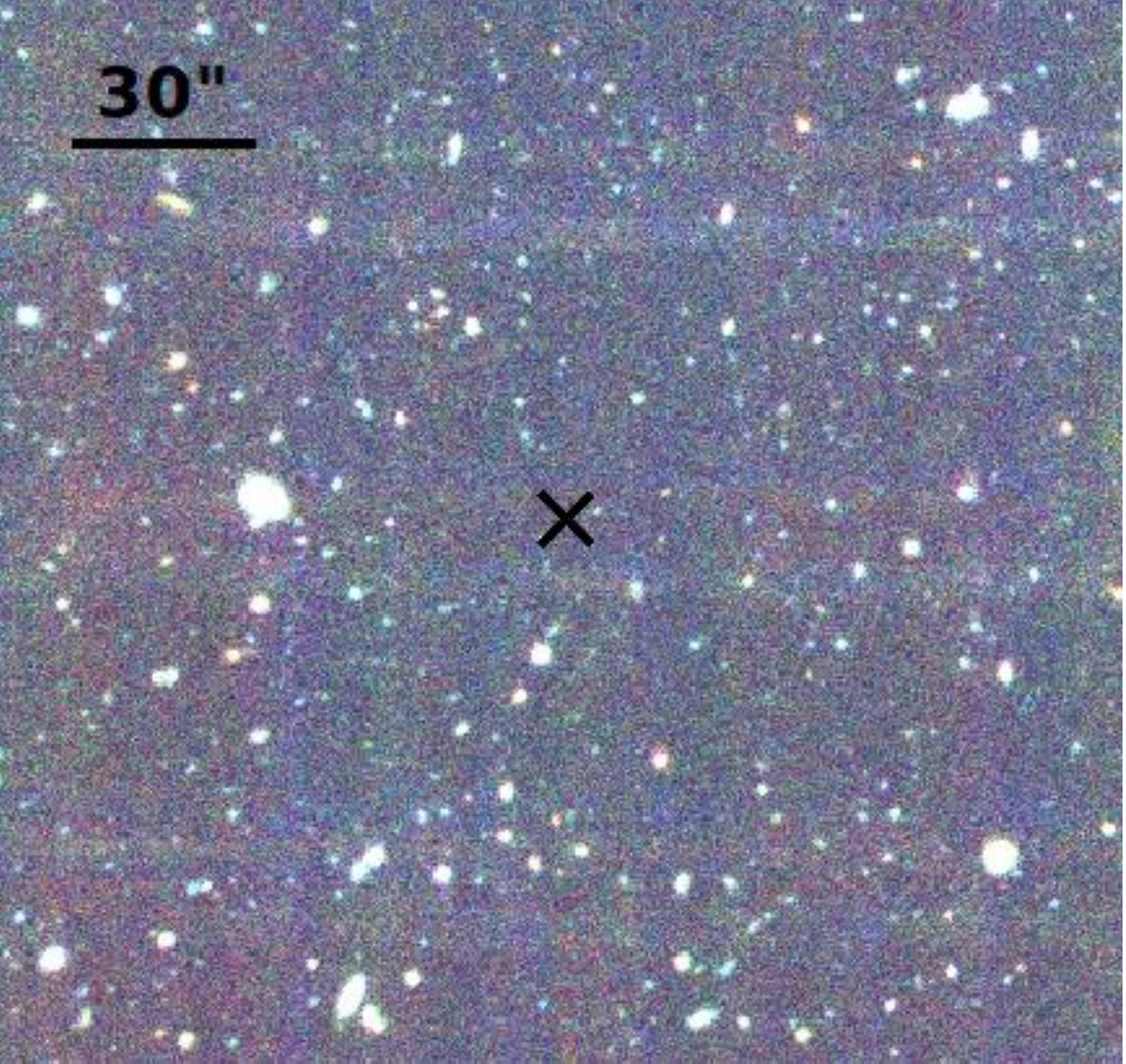} \\
{\bf d \hspace{2.2in}  e \hspace{2.2in} f }\\
\vspace{0.1in}
\includegraphics[width=5.6cm]{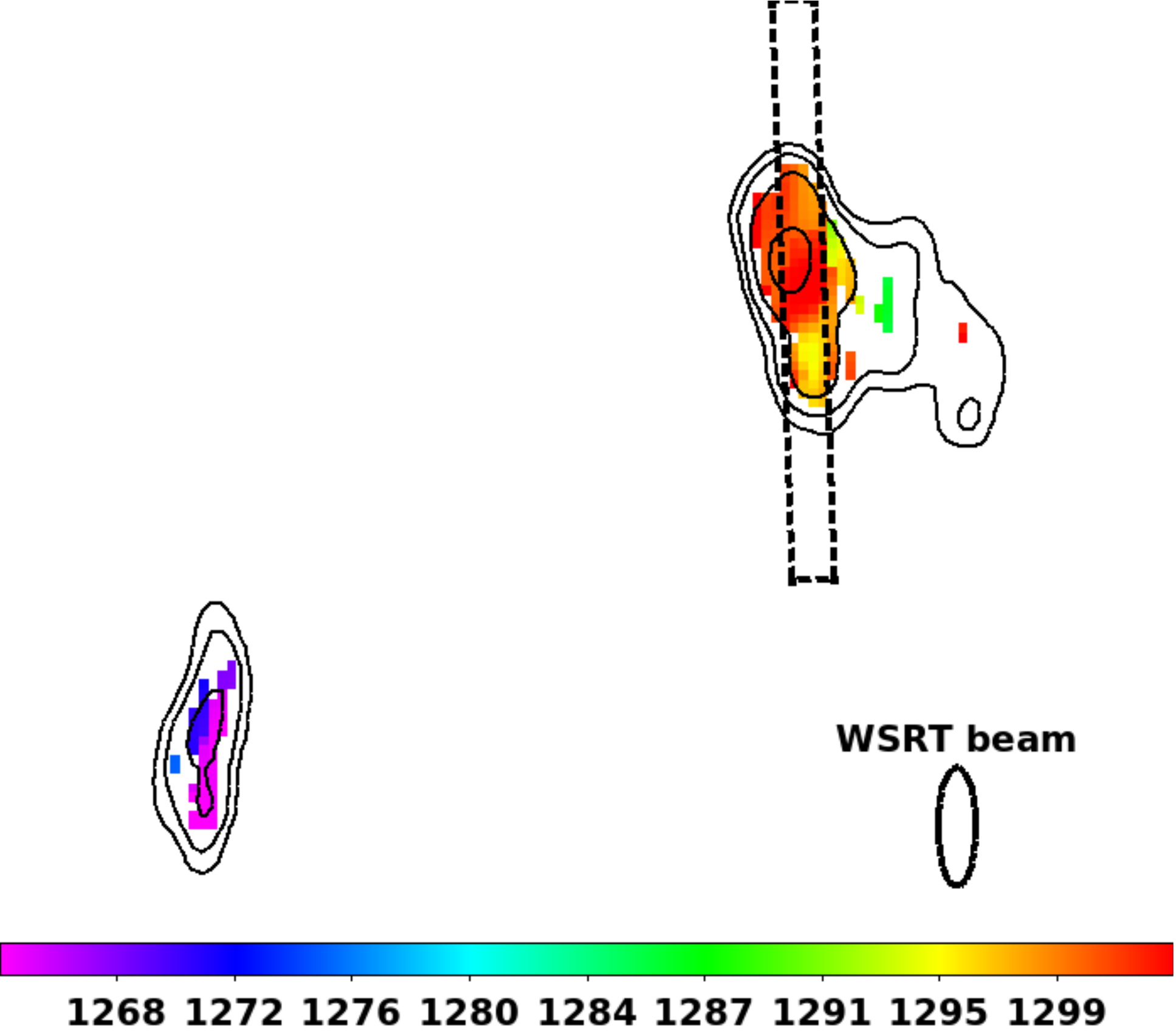} 
\includegraphics[width=5.5cm]{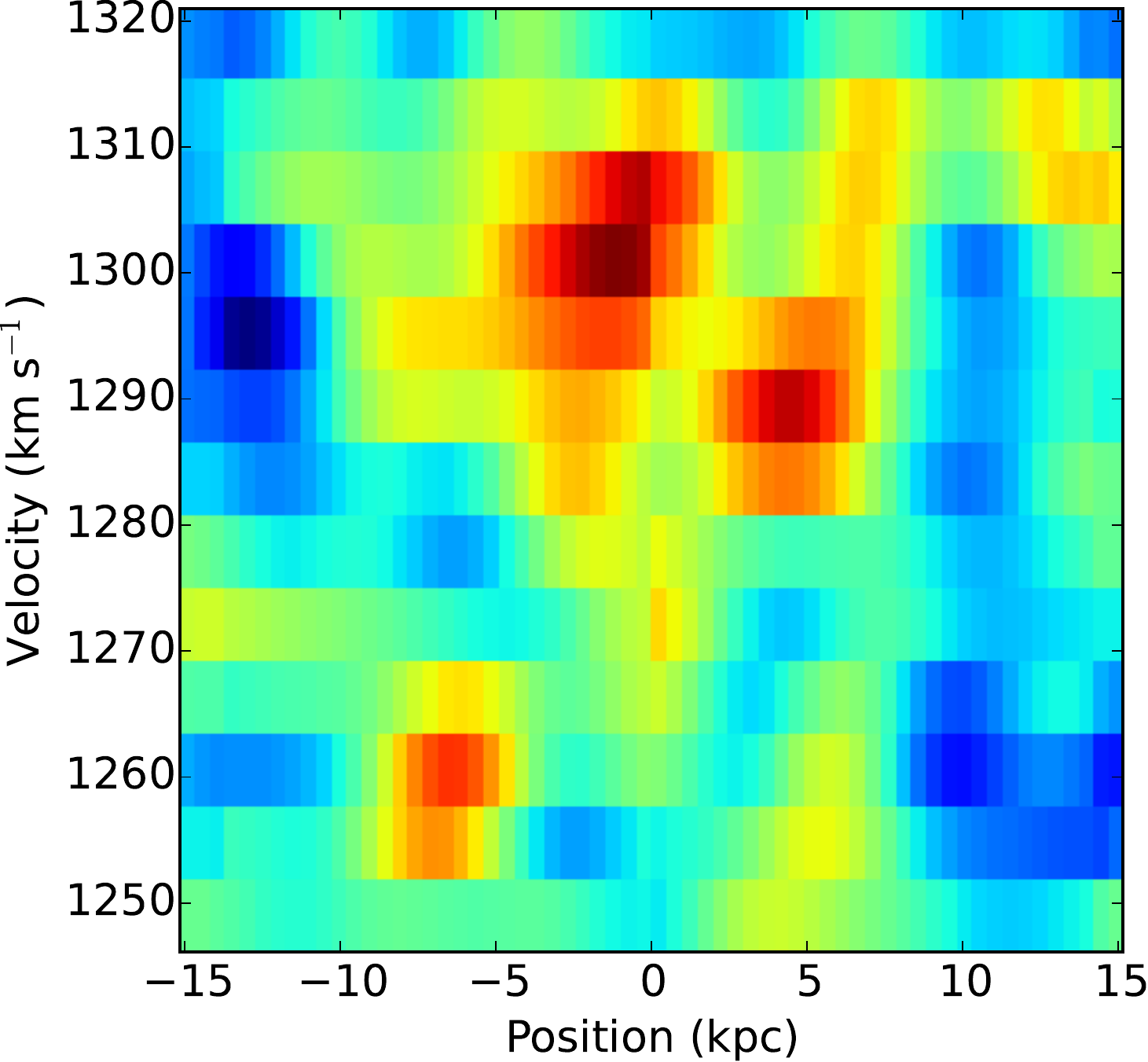}
\includegraphics[width=6.6cm]{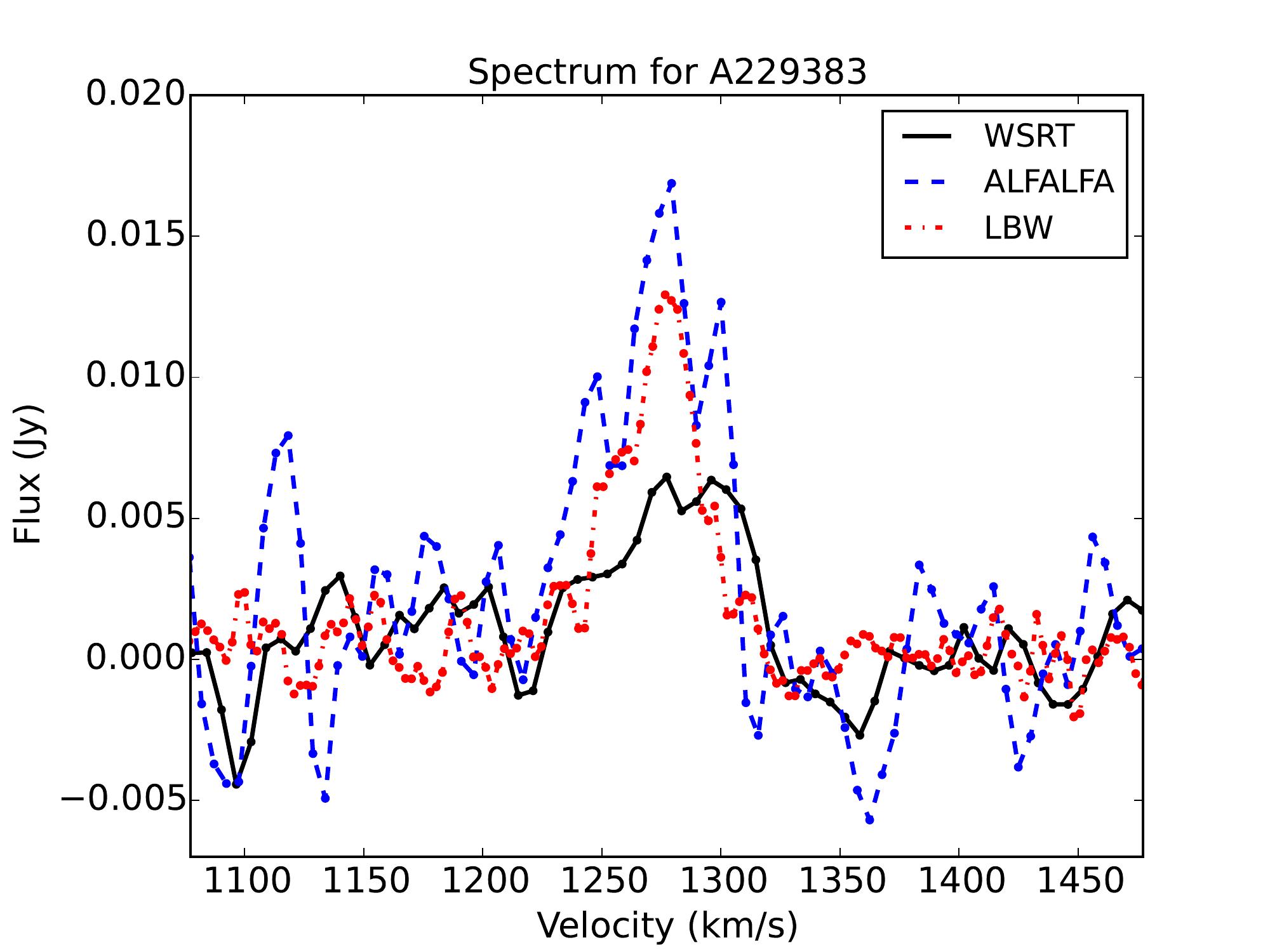}
\caption{
  Our observations of AGC~229383, same as Figure~\ref{385}.
{a)}~WIYN pODI optical image covering both peaks of the HI
   distribution, with contours from WSRT and black ``$\times$''s
   marking the HI peaks.
{b)}~A zoomed in view of the NW component, with a black
  ``$\times$'' marking the peak of the HI distribution. No optical
  counterpart is detected. An asteroid trail is also visible in this
  image.
{c)}~A zoomed in view of the unresolved SE component, with its
   center marked. Again no optical counterpart is detected.
{d)}~The moment 1 map from WSRT with the same HI contours
  as above. The ellipse in the bottom right corner indicates the size
  and shape of the WSRT beam. The dashed rectangular region shows the
  slice along the major axis which is used to generate the P-V
  diagram to the right.
{e)}~Position-Velocity diagram from WSRT. No ordered rotation
  is visible in this splotchy and irregular velocity field.
{f)}~ALFALFA, WSRT, and LBW HI spectra of AGC~229383. As
  discussed in the text, LBW recovers $64\%$ of the ALFALFA HI
  flux, and WSRT recovers only $52\%$. This likely means that there
  may be extended gas below the WSRT 
  sensitivity which connects these sources, and we consider them to be
  peaks of a common source in this work.
\label{383}
}
\end{figure*}

The HI velocity field of AGC~229385 shown in Figure~\ref{385}d is
equally difficult to  
interpret. The narrow {integrated} HI line width and single
peaked spectrum (shown 
in Figure~\ref{385}f) is suggestive of slow rotation. If we assume a
thermal velocity 
dispersion of $11.0$~km/s, the $34$~km/s {integrated} line
width of AGC~229385 
gives an observed rotation velocity of $16$ km/s when subtracting thermal
velocity and dividing by $2$. This translates to a rough
inclination-corrected rotation velocity of $18$ km/s. Indeed, the
moment~1 map shown in Figure~\ref{385}d shows evidence of
ordered rotation roughly along 
the major axis of the source, but the gradient is asymmetric/
The irregular shape of the P-V diagram for
AGC~229385 (Figure~\ref{385}e) 
further diagnoses this asymmetry. The southern side of the source
shows a clear slope of $12-15$ km/s, but then any gradient appears to
flatten out or even turn over as one approaches the north side of the
galaxy. Further, the P-V diagram reveals that the velocity dispersion
is of a similar order to the velocity gradient. It is
possible that the major axis 
of rotation is offset from the surface density major axis: fitting a
P-V diagram at $0^\circ$ removes any turnover in the north side of the
object, but does not give a significant gradient.

The FUV image of AGC~229385 from GALEX is shown in Figure
\ref{uv}. While this image is less striking than the
optical images, this source is still detected in both the NUV and FUV
images. The GALEX pipeline shreds 
this extended source into multiple point sources, so we measure its
brightness in the same $32''$ radius aperture and with the same
masking that was used on the optical 
images. The apparent magnitude and uncertainties in the NUV and FUV
bands are $19.631$~$(0.069)$~mag and $19.155$~$(0.035)$~mag,
respectively. After correcting for Galactic extinction, we use the
assumed distance of 25~Mpc to determine star formation rates (SFRs)
from the NUV and FUV luminosities following the relations of
\citet{murphy11} and \citet{hao11}, and report the results in
Table~\ref{abs}. We do not include a correction for internal
extinction.

Using the flow model distance of 25~Mpc, we derive
absolute global parameters for AGC~229385, which are listed in
Table~\ref{abs}. AGC~229385 has an optical luminosity comparable to
typical dwarf 
galaxies (converted to $M_B$$=$$-12.72$ via \citealt{jester05}). Its $g'-r'$
color is very blue, and corresponds to a $B-V$ color of $0.13$ (via 
similar conversion in \citealt{jester05}). Using this $B$~magnitude
instead of the SDSS $g'$~magnitude, we find $M_{HI}/L_B$$=$$38.2
M_\odot/L_\odot$. 
Using the self-consistent simple stellar population models discussed
in \citet{mcgsch14}, this color implies a stellar
mass-to-light ratio (in the $V$ band) of $M_\star/L_V$$=$$0.3
M_\odot/L_\odot$, 
or a total stellar mass of $\sim$$1.5 \times 10^6 M_\odot$, and a ratio
$f_{HI}$$=$$M_{HI} / M_\star$$=$$475$.


\subsection{AGC~229384} 
\label{AGC229384}

Figure~\ref{384}a shows the deep 3-color WIYN pODI
image of AGC~229384 with WSRT HI contours overlaid. No obvious optical
counterpart is visible. The faint grid of horizontal and vertical
stripes in the background is an artifact from the data reduction
process. Figure~\ref{384}b shows a zoomed-in view near the twin peaks
of the HI contours (marked with black $\times$s).  We used small $5''
\times 5''$ 
regions placed around this area to determine upper 
limits on the optical non-detection, using the same method as for
AGC~229385, after masking all obvious foreground and background
sources. The $3\sigma$ upper limits on this non-detection are 
given in Table~\ref{abs}, and are $27.9$, $27.7$, and $26.8$~\magsec
\, in $g'$, $r'$, and $i'$, respectively. To estimate an upper limit
on an integrated magnitude, we must assume an aperture size. Since the
HI major axis of AGC~229384 is $\sim$$50\%$ of the HI major axis of
AGC~229385, the aperture is scaled by the same amount, to
$15''$. Using 
this aperture of radius $15''$, we find $3\sigma$ upper
limits on the integrated magnitude in $g'$, $r'$, and $i'$ filters to
be 20.7, 20.5, and 19.7~mag, respectively. These upper limits
are used to generate upper limits on $M_{HI}/L_{g'}$, $M_\star$, and
$M_{HI}/M_\star$, all of which are shown in Table~\ref{abs}. In
particular, these non-detections correspond to a stellar mass upper
limit of $M_\star$$<$$3.4 \times 10^5 M_\odot$.
When necessary, the observed colors of AGC~229385 were used to make
filter conversions on the optical non-detections for AGC~229384.  An
archival GALEX image that covers AGC~229384 also shows no optical
counterpart for this object (see Figure \ref{uv}).

The HI contours of AGC~229384 in Figure~\ref{384}a show an irregular
distribution with two weak density peaks. The HI major axes at a
surface density of 1\msun~pc$^{-2}$ are $1.4' \times 1.2'$ or
$10$~kpc~$\times$~$8$~kpc. The HI velocity field shown in Figure
\ref{384}d is patchy and irregular,
and appears to be dominated by random motions. The position-velocity
diagram in Figure \ref{384}e shows no evidence of ordered rotation.
At its assumed distance of $25$~Mpc, AGC~229384 has a total HI mass of 
$M_{HI}$$=$$2.0 \times 10^8 M_\odot$. Complete details are given in
Tables~\ref{HIdat} and~\ref{abs}.

\subsection{AGC~229383}
\label{AGC229383}

From the ALFALFA HI observations, AGC~229383 was extracted as a
single, possibly extended weak source. Follow up observations with the
single pixel L-Band Wide (LBW) receiver confirmed the existence and
{extended nature of the
source,} since 
the more sensitive LBW observations only recovered $64$\% of the
original ALFALFA flux, as shown in Figure~\ref{383}f. WSRT observations
resolve the source into two low HI column density clumps, separated by
$5.5$ arcminutes 
{($\sim40$ kpc at $D=25$ Mpc)}.
The SE clump is only detected in a single beam 
at low signal to noise ratio, but is detected in both WSRT
pointings which overlap its position. It is possible that these two
clumps are independent sources. However, the two clumps together only
recover $52$\% of the original ALFALFA flux, suggesting that there may
be gas connecting the sources below the sensitivity of the synthesis
observations. {AGC~229383 may be two distinct sources, but
  because the missing flux and the clumpy HI distribution are
  ambiguous as to the true nature of the source, we choose to discuss
  it as a single source with two peaks in the remainder of this paper.}

Figure~\ref{383}a shows the color image made from the $g'$ and $r'$
observations of AGC~229383, which is located $19'$ to the NW of
AGC~229385. As with AGC~229384, no optical counterpart is
visible. Figure~\ref{383}b shows the zoomed-in region around the NW
peak of the HI distribution. Again we masked obvious foreground and
background sources and used regions around this area to
measure the background statistics and determine upper limits on the
optical non-detection using the same method as for AGC~229385. These
$3\sigma$ upper limits are given in Table~\ref{abs}, and are $27.8$
and $27.3$~\magsec, in $g'$ and $r'$, 
respectively. Similarly, we find upper limits on integrated magnitudes
in a $15''$ aperture of $20.7$ and $20.2$~mag in $g'$ and $r'$,
respectively. We use these in the same way as AGC~229384 to generate
upper limits for the derived quantities for AGC~229383, all of which
are listed in Table~\ref{abs}. Figure~\ref{383}c shows the
zoomed-in region around the unresolved SE peak of AGC~229383, and
reveals no optical counterpart as well.

The HI contours of AGC~229383 in Figure~\ref{383}a show the two
density peaks, of which the NW component appears resolved and the SE
component unresolved. The HI never reaches a projected HI surface
density of $1 M_\odot$pc$^{-2}$ (corresponding to a column density of
$12.5\times10^{19}$cm$^{-2}$ at 
this distance), and has a maximum extent of only $48''
\times 17''$ at a
column density of $5$$\times$$10$$^{19}$cm$^{-2}$. The HI velocity
field is patchy and irregular, and appears to be dominated by random
motions, though our data are limited due to the
fact that the two peaks are poorly resolved or unresolved. The limited
data and patchy nature of the velocity field precluded meaningful
fitting with XGAUFIT, so instead we show the moment 1 map in
Figure~\ref{383}d, and the messy position-velocity diagram in
Figure~\ref{383}e. At its  
assumed distance of 25~Mpc, AGC~229383 has a total HI mass of
$1.2\times 10^8 M_\odot$.

\subsection{Isolation, Environment, and Distance Uncertainty}

\subsubsection{Isolation of the HI1232+20 system}
\label{isolation}

Since many ``dark'' galaxy candidates turn out to be tidal features
(e.g., VIRGOHI21 \citet{duc08}) rather than isolated galaxies, we look
for possible objects which may have recently tidally interacted with
HI1232+20. To test this possibility, we searched all cataloged nearby
sources in the Arecibo General Catalog (AGC), NED, and the SDSS
spectroscopic survey and 
determined the timescale on which they
could have interacted with this system, given their current
velocities. If we generously assume that a flyby encounter may have
had a relative velocity of  $\sim$$500$~km/s, we can calculate 
how long ago the nearby objects could have interacted. Naturally, the
three sources in the HI1232+20 system all have short interaction time
scales with each other ($<$$300$ Myr, based on these assumptions). 
The only source with an interaction 
time scale $<$$1$~Gyr is AGC~742390, which is $\sim$$30'$~W of this
system and has $cz=1127$~km/s. AGC~742390 is an elongated star-forming
galaxy with  
$M_g = -15$~mag, assuming it is at the same distance of
$25$~Mpc. At this distance, AGC~742390 has a projected separation from
the HI1232+20 system of $\sim$$200$~kpc, but its smaller recession
velocity ($\Delta v \sim 200$ km/s) implies that it is likely more
nearby than this system.
Given the lack of obvious
nearby galaxies in optical and HI 
surveys, and the lack of objects which could have recently tidally
interacted with this system, it seems that HI1232+20 is a locally
isolated system, and not a tidal feature of a larger { parent
  object. Still, we cannot exclude the possibility that this system
  may have been produced as a result of tidal interactions or other
  gas stripping processes.}

\begin{figure}[tb]
\centering
\includegraphics[width=8.5cm]{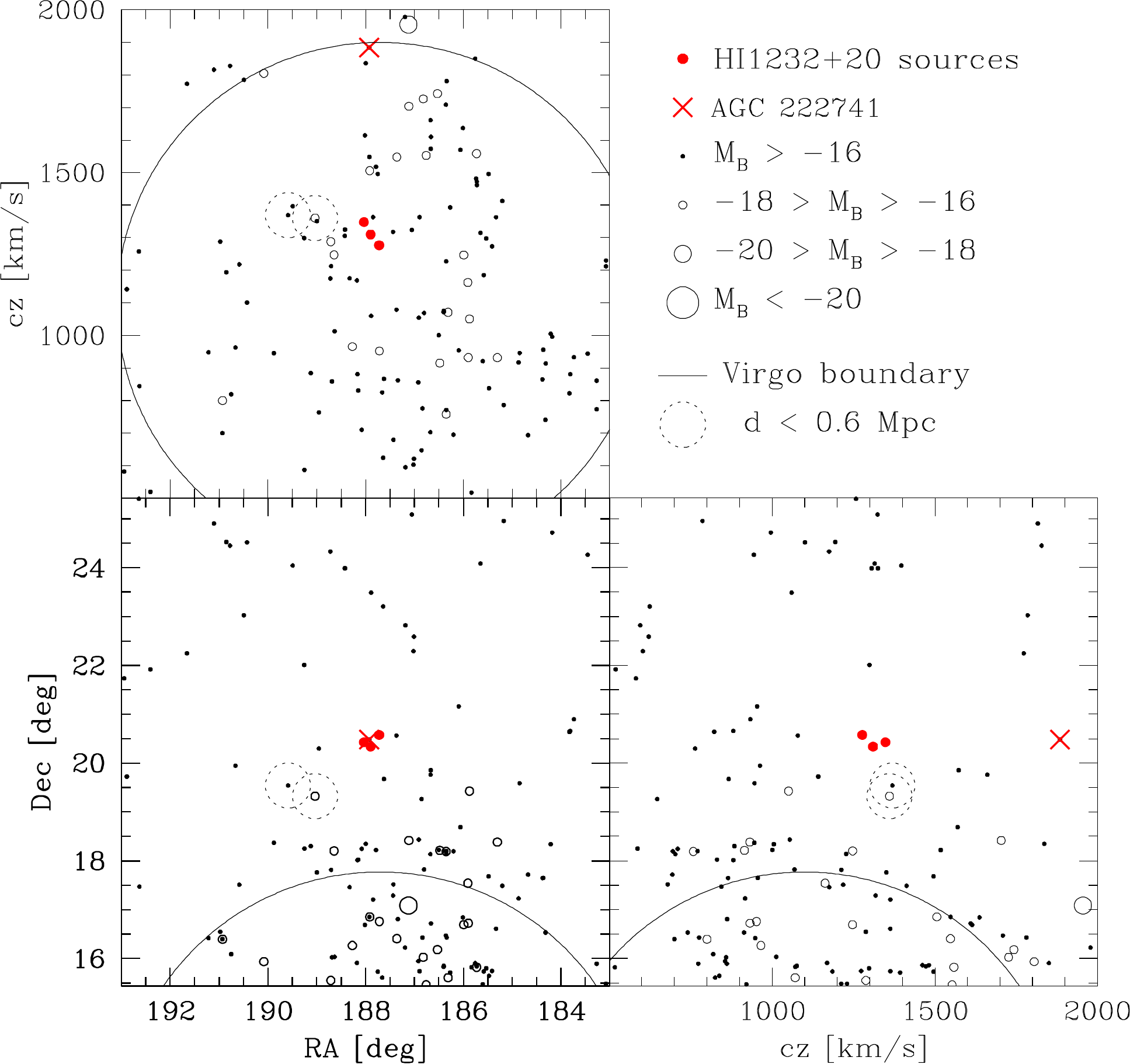}
\caption{
 All panels show galaxies from the UZC, SDSS, and AGC.
 Larger circles
 represent more luminous galaxies, and 
 smaller circles and dots represent less luminous galaxies.
 Note that we are plotting the three objects in the HI1232+20
 system as red
   dots, as well as the background spiral galaxy (AGC~222741) at
   $cz$$=$$1884$~km/s as a red ``$\times$''. The latter is nearby the
   objects in the 
   HI1232+20 system in the RA-Dec plane, but is well-separated in
   the other two panels. 
 The bottom left panel shows the galaxy distribution in the plane of
   the sky.
 The top left panel shows the galaxy distribution in the
   velocity-right ascension plane.
 The bottom right panel shows the galaxy distribution in the
   declination-velocity plane.
 The solid curved lines show the virial radius of the Virgo Cluster in
   each projection.
 Dotted circles enclose the galaxies nearest to this system.
   \label{env}
}
\end{figure}


\subsubsection{Effects of distance uncertainty and the environment}
\label{denv}

Throughout this work we have adopted a flow-model distance
\citep{masters05} of $D$$=$$25$~Mpc to the HI1232+20 system. However,
given its location on the outskirts of the Virgo Cluster, there is
some uncertainty about its true distance. We consider the possibility
that the HI recession velocity may not give an accurate distance for
this system, which could affect the derived absolute properties of the
objects in the HI1232+20 system.

Large scale peculiar motions have been observed around the Virgo
Cluster of galaxies, even beyond its virial radius,  
due to its significant gravitational influences. Recently,
\citet{kar14} used a large sample of 1801 galaxies (mostly from
\citet{karnas10} and with some new observations) in
the vicinity of the Virgo Cluster which have 
independent distance measurements (e.g., Tully-Fisher, TRGB, Cepheid)
as well as measured recession velocities. This kinematic sample of
galaxies was used to map
out the zero-velocity surface around the cluster, which encloses the
region of 
space where galaxies are falling into the Virgo
Cluster. \citet{kar14} find that the zero-velocity surface radius
is $7.2\pm0.7$~Mpc, which corresponds to a projected radius of
$25^\circ\pm2^\circ$ at their
 assumed Virgo distance of $D$$=$$17.0$~Mpc. The
HI1232+20 system is at a projected 
distance of only $\sim$$8^\circ$ from the center of the Virgo Cluster
(NGC~4486), and may be participating in the infall motion. 
Figure~1 in \citet{karnas10} shows a graphical
representation of the difficulty in determining distances from
recession velocities in this region, and for 
this system's measured recession velocity of $\sim$$1300$~km/s, there
are three possible distances. If HI1232+20 were infalling from the
near side, located at the center of the cluster, or infalling from the
far side, it could have distances of $\sim$$12$, $\sim$$17$, or
$\sim$$25$~Mpc, respectively. 
Further complicating the nearby velocity
field 
is the Coma~I cloud just north of the HI1232+20 system. This complex
of galaxies with peculiar velocities is centered around
$(\alpha,\delta) = (12.5\textrm{h}, +30^\circ)$ \citep{kar11}.
Still, we can make a crude but reliable
estimate of the lower limit on the distance based on the
fact that we do not resolve any individual stars in the optical
counterpart of AGC~229385. WIYN has been used 
to successfully resolve stellar populations in
galaxies under similar observing conditions out to $2$$-$$4$~Mpc
(e.g., Leo~P, \citealt{rhode13}; M81 group, Rhode, private
communication). Furthermore, WIYN observations of SHIELD galaxies
\citep{cannon11} resolve upper main sequence and supergiant stars at
distance of $8$~Mpc.

Many of the most extreme properties of the objects in the HI1232+20
system are distance-independent quantities and would not be affected
by a more 
nearby distance (e.g., $M_{HI}/L_B$, surface brightness measurements
and limits, average HI surface density). However, if they were only
$\sim$$12$~Mpc distant instead of $25$~Mpc, the absolute quantities
(e.g., HI mass, stellar mass, total 
luminosity, physical area) would all scale  down by a factor of
four. For example, the HI mass of AGC~229385 would become $1.8 \times
10^8 M_\odot$, its stellar mass would be reduced to $3.8 \times 10^5$
$M_\odot$, its absolute B~magnitude would be reduced to
$M_B$$=$$-11.4$~mag, and its optical diameter would be
$\sim$$2$~kpc.

Even with the difficulties of constraining the absolute distance to
the HI1232+20 system, we are interested in the large scale environment
around it, and how isolated it has been on longer time scales. The
background spiral galaxy AGC~222741 (CGCG~129-006, labeled in
Figure~\ref{allim}) 
appears on the sky between the three sources in HI1232+20, but has a
recession velocity of $v_{\rm lsr}$$=$$1884$ km/s, which is substantially
higher than the velocities of the three sources in the system ($1277$,
$1309$, and $1348$~km/s), so is very unlikely to be related. 
Figure~\ref{env} shows the galaxies in the area around these HI sources.
Galaxies shown on the plot come from the Updated Zwicky Catalog (UZC,
\citealt{falco99}), from the spectroscopic sample of SDSS DR9
\citep{ahn12}, and from the Arecibo General Catalog (AGC).
The UZC is an extragalactic redshift survey of 
$\sim$$20,000$ galaxies that is $96\%$ complete to
$m_{Zw}$$<=$$15.5$~mag
and the SDSS DR9 spectroscopic
survey includes spectra of $\sim$$1.5$ million galaxies, and is $95\%$
complete down to $r'$$=$$17.7$~mag.
On Figure~\ref{env}, a red ``$\times$''
indicates the location of the background spiral galaxy (AGC~222741),
which appears near our HI sources on the sky.
Also shown on Figure~\ref{env} is the projected virial radius of the
northern subcluster of 
the Virgo Cluster of galaxies \citep{binggeli85}. The curves
enclose a region centered at the position of M87 with a radius of
$5.4^\circ$ and 
centered on a recession velocity of $1100$ km/s and extending
$800$ km/s on either side \citep{ferrarese12}.

In order to demonstrate the complexity of the velocity field around
HI1232+20, we calculate 3-space separations between this system and any
nearby sources, using only their positions and observed
velocities. The three sources within the HI1232+20 system are
all within $\sim$$20'$ and $\sim$$70$~km/s of each other. While
the global velocity field around these sources is
complicated due to the influence of the nearby Virgo Cluster
\citep{karnas10, kar14}, 
we use this approach to crudely identify any possible galaxies which
may be near enough to affect this system, and have similar
positions and velocities. The object with the most similar velocity
and position is NGC~4561 with a velocity of $1360$~km/s and 
an angular separation of $1.4^\circ$, implying a physical separation
of $\sim$$400$~kpc assuming a simple Hubble flow. 
However, the Tully-Fisher distance to NGC~4561 is $12.3$~Mpc
\citep{tf87}, so it is likely infalling to the Virgo Cluster from the
near side.
The object with the next-smallest 3-space distance is the starbursting
galaxy IC~3605, which has a velocity of 
$1360$~km/s and is located $\sim$$1.8^\circ$ to the SE, with an
implied physical separation (Hubble flow only) of $\sim$$500$~kpc. No
other distance measurements exist for IC~3605.
The locations of NGC~4561 and IC~3605
are indicated on Figure~\ref{env} with large dotted circles. The close
proximity of this system to the Virgo Cluster means that its distance
is uncertain and that the global velocity field is rich and complex.

\section{Discussion}\label{discussion}

The objects in the HI1232+20 system are not easily explained, and some
of their properties seem contradictory and puzzling. {For
  example,} it is difficult to {understand} the star formation
history of AGC~229385, which 
apparently has only produced a tiny population of stars in an
otherwise massive HI cloud. {The HI mass of AGC~229385 (log
  $M_{HI} = 8.9$) is larger than the nearby star-forming Large
  Magellanic Cloud (log $M_{HI} = 8.7$, \citealt{kim98}) and just 
  smaller than the nearby spiral galaxy M33 (log $M_{HI} = 9.1$,
  \citealt{gratier10}),
  although its stellar populations and optical luminosity are vastly
  dissimilar.}
The HI kinematics of its HI cloud are
also perplexing, as its rotation {speed} seems inadequate for
its large mass 
and size. It is similarly difficult to explain why the other members
of the HI1232+20 system have not formed any detectable stars, even
with their substantial, although quite spread out, HI
distributions. {For comparison, both AGC~229383 and AGC~229384
  have larger HI masses than the nearby dI galaxies IC~10 and NGC~6822
  (log $M_{HI} = 8.0$, \citealt{nidever13}, \citealt{deblok2000}), but lack any optical
  counterparts in our observations.}

In order to put these objects in context with
other galaxies, we consider their locations in typical galaxy scaling
relations. {In this exercise, we are treating these objects as
  independent galaxies and not simply gas clouds that have been
  stripped or tidally 
  disturbed.} It is possible that objects like this may be part of a
large but mostly-unobserved class of galaxies (e.g., the sunken
galaxies of \citet{disney76}), but is more likely that the objects in
this system are simply unique and uncommon galaxies. 
Since AGC~229385 has an optical counterpart we consider
its location on optical and HI scaling relations, while for the other
sources our optical upper limits can still help constrain some of the
same scaling relations. After discussing these scaling relations, we
will will consider some possible formation scenarios to explain this
unusual system of objects.

\begin{figure}[tb]
\centering
\includegraphics[width=8.6cm]{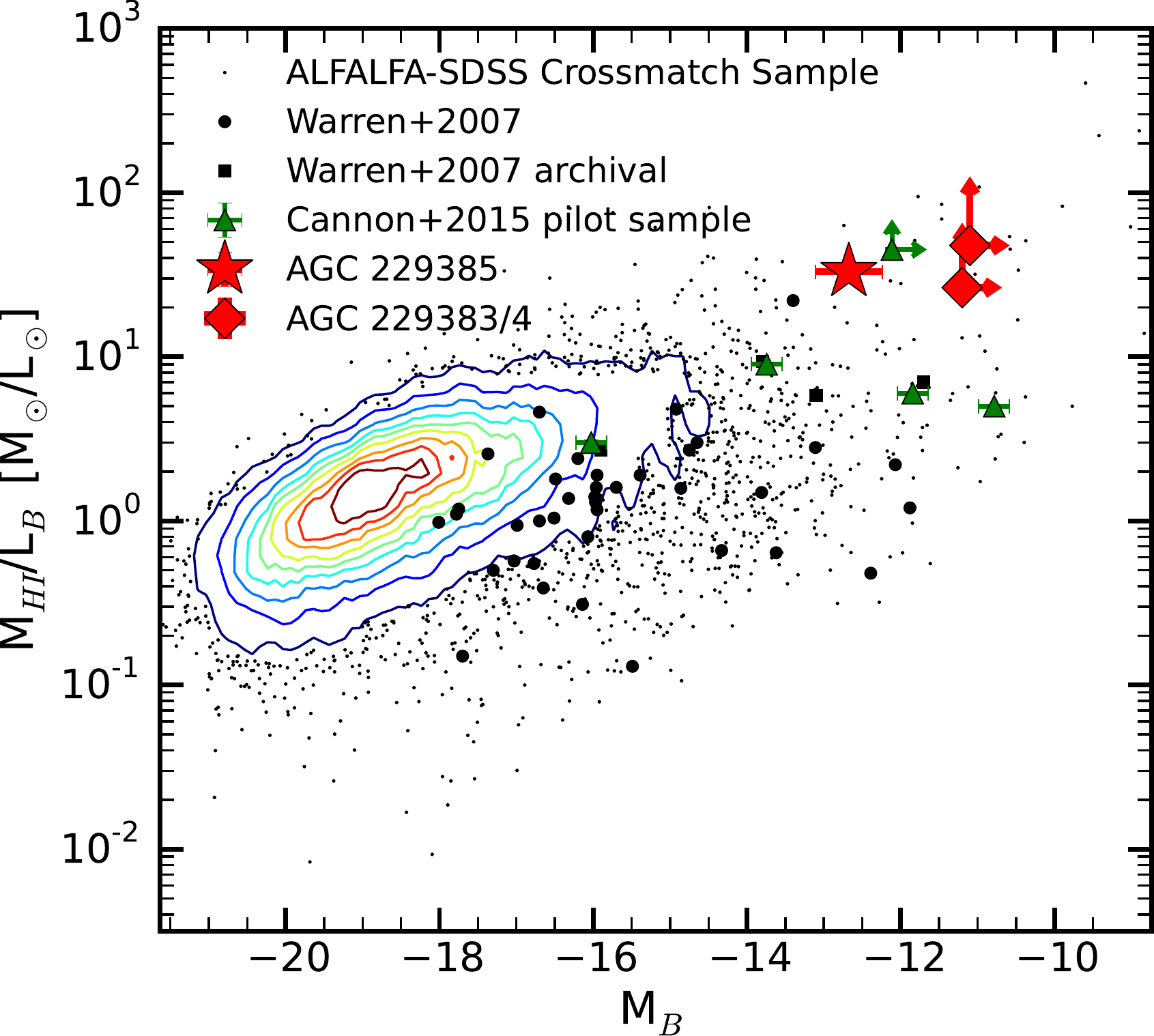}
\caption{Relationship between the HI mass-to-light ratio
  ($M_{HI}/L_B$, in solar units) and absolute $B$~magnitude. The small
  dots and contours indicate HI-detected galaxies from the ALFALFA
  $\alpha.40$ catalog matched with SDSS DR7 photometry. The dark
  circles and squares come from the new and archival
  observations of \citet{warren07}. The green points show the five
  (almost) dark galaxies from the pilot observations
  of \citet{cannon15}. AGC~229385 is shown as a red star, and the upper
  limits for AGC~229383 and AGC~229384 are indicated with diamond
  points and arrows. The error bars on AGC~229385 also include a
  distance uncertainty of $\pm5$~Mpc. The object from the
  \citet{cannon15} sample with a larger lower limit on $M_{HI}/L_B$
  than the measured value for AGC~229385 is AGC~208602 and is likely a
  tidal feature and not an isolated galaxy.
\label{ML} 
}
\end{figure}

\subsection{$M_{HI}/L$ relationship}

One of the most extreme properties of the objects in the HI1232+20
system is their exceptionally large HI mass-to-light ratio
measurements (or lower limits). Galaxies are known to follow a 
typical relationship between between this HI mass-to-light ratio and
the overall luminosity. The HI
mass-to-light ratio is defined as $M_{HI} / L$, where $L$ is the
optical luminosity, often measured in a blue filter. 
This relationship
is especially difficult to measure for faint low-mass galaxies, where
a significant fraction of the optical luminosity may come from low
surface brightness regions. Almost universally, whenever a galaxy with
a reportedly large $M_{HI}/L_B$ ratio is observed with deeper optical
images, the ratio returns to more typical values near
unity. \citet{warren04} used survey and catalog data to identify
possible galaxies with a large $M_{HI}/L_B$ ratio
($3$$<$$M_{HI}/L_B$$<$$27$), but their sample of 9 large $M_{HI}/L_B$
galaxies were almost 
all found to have less extreme ratios ($M_{HI}/L_B$$<$$5$) after deeper
observations. Similarly, \citep{vanzee97} used broadband optical
imaging observations of six low 
surface brightness dwarf galaxies to show that their catalogued
optical magnitudes had been severely underestimated by
$\sim$$1.5$~mag, so their previously reported $M_{HI}/L_B$ ratios
became $4$ times smaller and less extreme. 
Among dwarf galaxies, typical
measurements of $M_{HI}/L_B$ are between $0.15$ and $4.2$, considering
a variety of samples including Sm/Im galaxies
(\citealt{robertshaynes94}; \citealt{stilisrael02}) and field dIs
\citep{lee03}. The relatively small dynamic range of ratios (a factor
of $30$ between the lowest and highest ratio) highlights the
importance of careful and accurate measurements of this ratio.

Figure~\ref{ML} shows the relationship between the HI mass-to-light
ratio ($M_{HI} / L_B$) and absolute $B$~magnitude ($M_B$) for multiple
samples of galaxies. The main dataset shown in Figure~\ref{ML} comes
from the matched ALFALFA $\alpha.40$ and SDSS catalogs of \citet{haynes11},
which is the parent sample from which the ALFALFA (Almost) Dark
Galaxies are drawn. We note that the shallow SDSS photometry may
underestimate the 
luminosity of the faint sources in this sample. The average exposure
time in SDSS is only $\sim$$1$ minute, and will not detect low surface
brightness emission from these galaxies. {Additionally, optical
  fluxes may be erroneously estimated due to issues with SDSS
  background subtraction and effects from nearby bright stars.} 
 Also shown on Figure~\ref{ML}
are galaxies from \citet{warren07}, who observed 38 galaxies at
optical and radio wavelengths, and also compiled an archival sample of
previously observed galaxies. Some of these sources are observed at
low Galactic latitude, which results in additional uncertainty
in the necessary extinction corrections. The five (almost) dark
galaxies from 
the pilot VLA observations of \citet{cannon15} are also shown on
Figure~\ref{ML} as green dots. The VLA observations allowed OCs to be
identified for all but one of their sample, and the gas mass to
light ratios for these sources are well-measured.
We find that AGC~229385 has
$M_{HI}/L_{g'}$$=$$45.8
M_\odot/L_\odot$, or, converted to $B$ via \citet{jester05},
$M_{HI}/L_B$$=$$38.2 M_\odot/L_\odot$, and its position is indicated on
Figure~\ref{ML}. As AGC~229383 and AGC~229384 are not detected in our 
deep optical images, we can only determine lower limits on
$M_{HI}/L_{g'}$, and find $>$$31$ and $>$$57~M_\odot/L_\odot$,
respectively. These limits are also shown in Figure~\ref{ML}, where we
have converted our upper limits in $M_{g'}$ to $M_B$ assuming the same
$g'-r'$ color as AGC~229385.

\citet{warren07} suggest that there is an upper envelope in
Figure~\ref{ML}, which may represent the minimum amount of stars a
galaxy will form, given a shallow potential well and an isolated
environment. The sources in {the HI1232+20 system} are in an
extreme region of Figure~\ref{ML}, near this upper envelope. 
While there are other 
galaxies from ALFALFA with more extreme {values of $M_{HI}/L_B$
  shown on the plot, none have ratios that are} as well-determined as
the objects in this system. SDSS photometry 
for faint low surface brightness galaxies will likely underestimate
their luminosity, which leads to an overestimate of the 
$M_{HI}/L_B$ ratio. {A recent study identified a low surface
  brightness galaxy near 
  the Virgo Cluster with a very large HI mass-to-light ratio that they
  measure as $M_{HI}/L_V \gtrsim 20 M_\odot/L_\odot$
  \citep{Bellazzini15}. } 
AGC~229385 has the largest 
{accurately} measured HI
mass-to-light ratio in the literature, but still appears to lie along
a continuation  of the trend seen in more luminous galaxies.
 \citet{warren07}
use a similar sample of galaxies with HI and optical data
to fit the relationship between $M_{HI}/L_B$ and $M_B$. They fit the
upper envelope of the relationship with 
the following expression:

log$(M_{HI}/L_B)_{\rm max}$$=$$0.19(M_B + 20.4)$.

\noindent
For AGC~229385 its $M_B$$=$$-12.72$ would predict a maximum
$M_{HI}/L_B$$=$$29$. While AGC~229385 does follow the general trend of
low stellar mass objects having higher gas mass-to-light ratios, we
measure $M_{HI}/L_B$$=$$38$, which is even 
more extreme than the upper envelope of \citet{warren07}.

\subsection{Galaxy Scaling Relations with Stellar Mass}

\begin{figure}[htb]
\centering
\includegraphics[width=8.5cm]{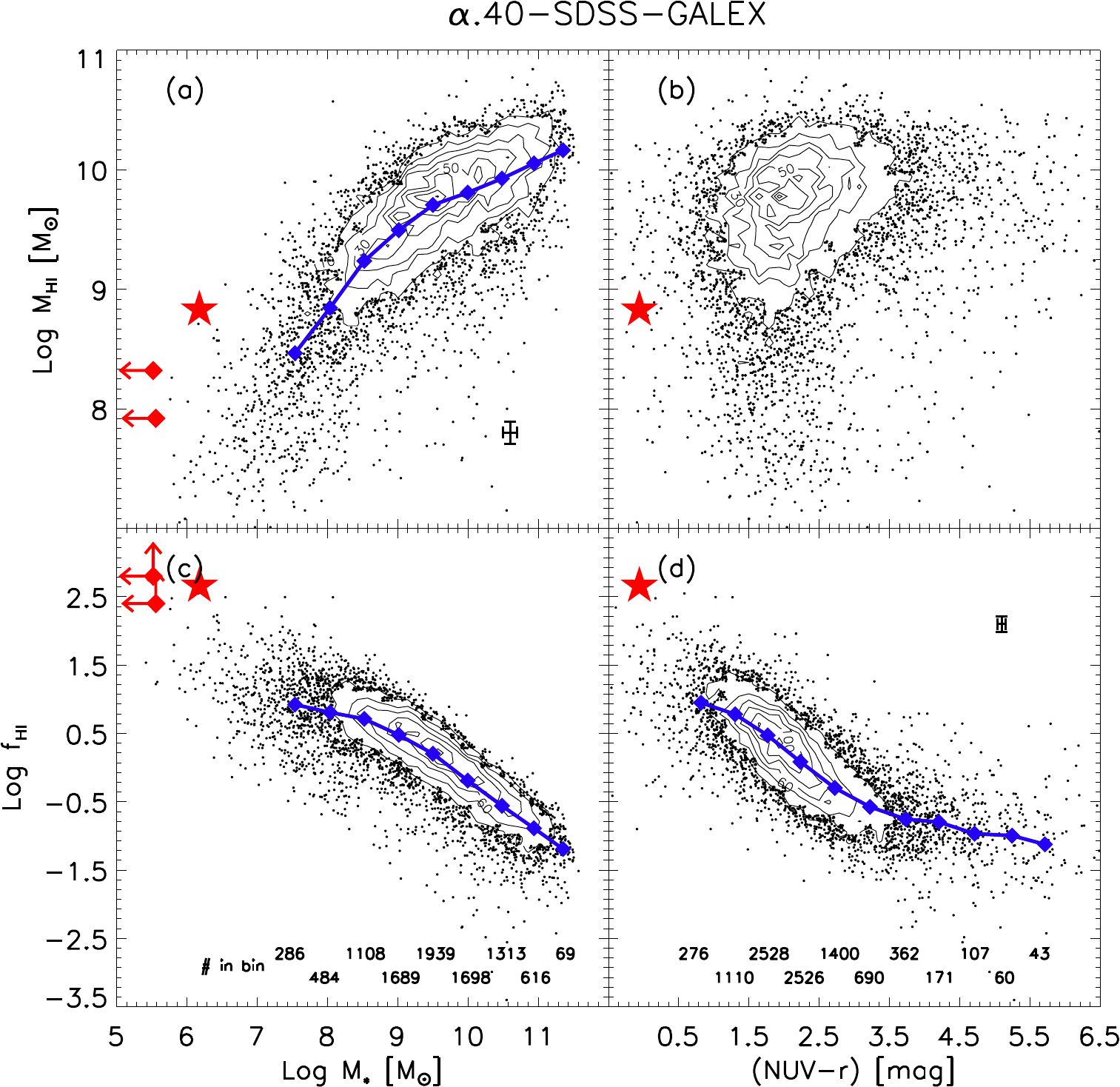}
\caption{The scaling relations between HI and optical observations,
  using the $\alpha$.40-SDSS-GALEX sample and analysis from
  Figure 8 of \citet{huang12}. In 
  all panels, number densities of galaxies are calculated 
  within grid cells set by the intervals of the minor ticks on the
  axes. Contours are drawn above densities of 20 galaxies per grid
  cell, and selected contours are labeled with their number
  densities. Blue diamonds and lines show the average y-values in each
  x-bin.
Panel (a) shows the relationship between HI mass and stellar mass,
  with stellar masses from mass-to-light ratios from fits to SDSS
  photometry. 
Panel (b) shows the HI mass as a function of NUV-r color.
Panel (c) shows the ratio of HI mass to stellar mass
($f_{HI}$$=$$M_{HI}/M_\star$).
Panel (d) shows the relationship between $f_{HI}$ to NUV-r color.
Typical error bars of individual galaxies are shown in the
  corner of panels (a) and (d).
The red star shows
  the location of AGC~229385 in each panel, while the red arrows
  indicate the upper limits of AGC~229384 and AGC~229383 where measured.
\label{huang}
}
\end{figure}

Studies of large samples of galaxies have found that stellar mass
seems to be an important parameter that relates to star formation in
the possible evolution of galaxies from the blue cloud to the red
sequence (\citealt{brinchmann04}; \citealt{salim07};
\citealt{huang12}). Using a sample of 9,417 ALFALFA-selected galaxies
with counterparts in archival GALEX and SDSS images, \citet{huang12}
studied the scaling relations as a function of stellar mass and
optical color (see their Figure 8). Figures~\ref{huang}a
and \ref{huang}c show the HI mass and $f_{HI}$$=$$M_{HI}/M_\star$ as a 
function of the stellar mass, which is determined from SED fits. A
clear relationship with $M_{HI}$ is found from
$M_\star$$=$$3.2$$\times$$10^7 M_\odot$ to $3.2$$\times$$10^{11}
M_\odot$, with a change in slope at 
$M_\star$$\sim$$10^9 M_\odot$. Analogously, $f_{HI}$ follows the same
general trend with a break at $\sim$$10^9 M_\odot$. On both panels the
location of AGC~229385 is indicated with a large star, and the upper
limit measurements 
of AGC~229383 and AGC~229384 are indicated with arrows. These sources
are deviant from the expected scaling relations at the low stellar
mass end, and have too much HI for their stellar mass (detected or
not).

Figures~\ref{huang}b and \ref{huang}d show $M_{HI}$ and $f_{HI}$ as a
function of $NUV-r$ color, which acts as an indicator of the amount
of recent star formation (UV) compared to the amount of past star
formation ($r$). Here only AGC~229385 can be plotted, since we have
no optical or UV detections of the other two sources. AGC~229385 is at
an extreme location in both of these parameter spaces, and lies
significantly above an extrapolation of the low mass trend in the
relationship between $f_{HI}$ and $NUV - r$. The color cannot be
much bluer than it already is, since after $5$~Myr, a simple stellar
population of half solar metallicity will have $NUV-r$$=$$-0.2$
\citep{bressan12}. AGC~229385 seems to be deviant in the sense that it
has too much HI for its $NUV-r$ color, consistent with the previously
discussed panels. Finally, we note that \citet{toribio11} found a
strong relationship between $M_{HI}$ and the optical diameter measured
at $\mu = 25$\magsec, such that galaxies with smaller values
$D_{25}$ typically had smaller HI masses as well. However, we cannot
compare the properties of the HI1232+20 system with this relationship
because the optical counterpart to AGC~229385 never reaches $\mu =
25$\magsec, and the other two sources have even fainter limits on
their optical non-detections.

\subsection{Galaxy Scaling Relations with HI mass}

We now discuss the scaling relations which depend on the total HI
mass of a galaxy. First, we consider the relationship between HI
diameter (measured at a particular
the column density) and total HI mass. \citeauthor{BR97}
(\citeyear{BR97}, hereafter BR97)  observed this 
relationship in a sample of  
108 spiral and irregular galaxies and found a strong correlation
between $log M_{HI}$ and $log D_{HI}$ (the HI diameter at $1
M_\odot/$pc$^2$), with a dispersion of only 
$0.13$ dex, and a slope of $1.96 \pm 0.04$. 
\citet{swaters02} found the same correlation and slope when they used
a sample of lower mass irregular galaxies. This relationship has been
measured between HI masses of $6 \times 10^7$ and $3 \times
10^{10} M_\odot$ and between HI diameters of $0.8$ and $160$~kpc. A
consistent relationship implies that there is a constant average HI
density in all gas-rich galaxies, which BR97 estimates as $3.8 \pm 1.1
M_\odot  $pc$^{-2}$.

While explanations for the underlying mechanisms for this correlation
are complex (e.g., feedback, turbulence, etc.), it is simple to apply
the relationship from BR97 to our sources. In 
the case of AGC~229385, our HI mass predicts an HI diameter of
$16$~kpc (measured at $1 M_\odot/$pc$^2$). Our HI observations of
AGC~229385 show an elongated source with dimensions $24 \times
10$~kpc, when measured at this HI column density. This is $\sim$$50\%$
larger than predicted, and while projection effects could reduce this
discrepancy somewhat, the HI kinematics do not suggest we are viewing
a disk edge-on so any correction for projection effects will be
small. AGC~229384 has an HI mass 
of $2.00 \times 10^8 M_\odot$ which predicts an HI diameter of
$8.1$~kpc. Our observations show an elongated HI source with 
dimensions $10 \times 8$~kpc, which is slightly larger than expected
(although this source has two separate HI peaks at that column
density). The HI distribution of AGC~229383 never reaches the column
density that corresponds to $1 M_\odot/$pc$^2$, and we are unable to
apply this relationship to it. Both AGC~229385 and AGC~229384 have
larger HI diameters than predicted by the HI diameter-HI mass scaling
relation. This also implies that their 
average HI surface density is significantly less than the constant
value found by BR97. 
Assuming a circular HI disk for AGC~229385 and AGC~229384, we find that
their average HI surface densities are $1.6$
and $2.4$$M_\odot$pc$^{-2}$, 
respectively, which are at the lower limit of the 
distribution found by BR97. \citet{rosenbergschneider03} also compared
$M_{HI}$ and $D_{HI}$ for a different sample, but their relationship
gives similar results to BR97. These unusually low HI surface densities
may be related to the lack of significant star formation in the
objects in the HI1232+20 system.

\subsection{Galaxy Scaling Relations with HI kinematics}

We consider the HI kinematics of the sources in the HI1232+20 system,
especially with regard to their apparently slow rotation. However,
since the HI rotation curves of these sources are difficult to fit or
interpret (see Figures  \ref{385}e, \ref{384}e, and \ref{383}e), we
instead use the {integrated} width of the 21cm line itself to
measure their 
rotation. The HI velocity widths ($W_{50}$, measured at $50\%$ of
the peak value) of the three sources in the HI1232+20 system are
unusually small for their HI masses. We model the HI velocity width
distribution for all ALFALFA galaxies as a function of their HI
mass. Integrating over the model distribution at the HI mass of
AGC~229385 ($M_{HI}$$=$$6.7 \times 10^8 M_\odot$) we find that only
$2\%$ of ALFALFA galaxies have velocity 
widths smaller than its measured value of $W_{50}$$=$$34$ km/s. Similarly
for AGC~229384, we find only $3\%$ of ALFALFA galaxies at its HI mass
have similarly small values of $W_{50}$. We note that the much wider velocity
width of AGC~229383 falls near the 50th percentile of objects of its
HI mass, but that this may be due to the presence of multiple objects
within the ALFALFA beam. Full details of the velocity width model will
be published in a forthcoming paper (Jones \etal in prep.). 

\begin{figure}[tb]
\centering
\includegraphics[width=8.5cm]{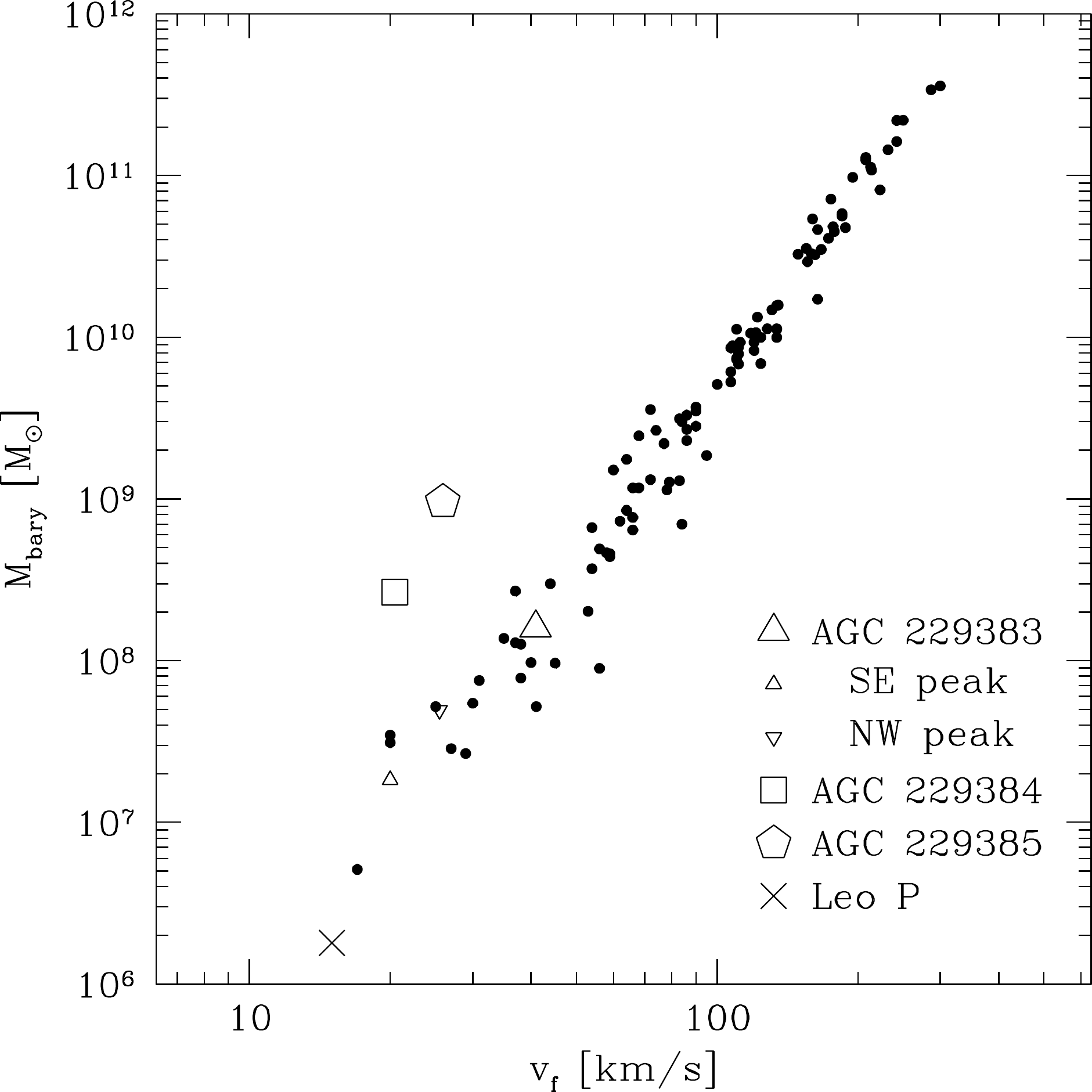}
\caption{
Baryonic Tully-Fisher relation. Black dots show galaxies
  measured by \citet{mcgaugh12} and \citet{mcgaugh05}. Open polygons
  show the sources in the HI1232+20 system from this work. AGC~229385
  (shown as an open pentagon) has the largest HI 
  mass of this sample and is offset above the BTF relation. The next
  most massive, AGC~229384 (shown as an open square), is also offset
  above the relationship, 
  while the least massive, AGC~229383 (shown as an open triangle), is
  consistent with the BTF 
  relation. {The two HI peaks of AGC~229383 are measured
    separately (shown as small triangles), and both lie along the BTF
    relation as well.} The ``$\times$'' shows the position of Leo~P, a
  metal-poor 
  gas-rich dwarf galaxy  discovered by ALFALFA
  in the local volume just outside the Local Group { 
  \citep{giovanelli13, rhode13, berncoop14}. }
 \label{btf} 
}
\end{figure}

We can next use the Baryonic Tully Fisher relation \citep{mcgaugh12}
to compare the total 
baryonic masses with the rotation velocities for the members of the
HI1232+20 system. 
Since these objects are gas-dominated, we use the
relationship from \citet{mcgaugh12} to calculate rotation velocities
as $v_f$$=$$W_{20}/2$, where $W_{20}$ is the HI velocity width at
$20$\% of the maximum. {While the HI rotation of these objects
  is difficult to measure accurately,
  their placement on the
  BTF relation is intended as a suggestive exercise to shed light on
  some of their unusual properties. If these objects are indeed
  galaxies, then their rotation seems too slow for their measured
  mass.}

Figure~\ref{btf} shows the
locations of the objects in the HI1232+20 system compared to a
large sample of galaxies 
(\citealt{mcgaugh12}, \citealt{mcgaugh05}). The baryonic mass is the
sum of the 
stellar and total gas mass, and in gas-dominated galaxies without
substantial stellar populations, is determined as
$M_{\rm bary}$$=$$1.33 \times M_{HI}$, where the extra factor of
$1.33$ accounts for 
helium. The stellar mass of the optical counterpart of AGC~229385
contributes only a negligible $\sim$$0.2\%$ compared to its HI
mass. The upper 
limits on the non-detections of stellar populations of the other two
objects indicate that these would contribute similarly negligible
amounts of baryonic 
mass. 
This estimate of baryonic mass does not include
contributions from 
molecular or ionized gas. It is conceivable that there may be an
envelope of low density ionized gas surrounding these galaxies,
contributing more mass than is included in our determination of
$M_{\rm bary}$. 

Figure~\ref{btf} shows that the two more massive objects in the
HI1232+20 system both fall significantly above the standard BTF
relationship, while the least massive (AGC~229383) is in
agreement. {However, since AGC~229383 has two strong HI peaks
  which are separated by $\sim 40$~kpc, we also consider the
  kinematics of each clump separately, and plot them on
  Figure~\ref{btf} as well. AGC~229383 is the lowest column density
  object of this system, and is, at best, difficult to
  interpret. Naive placement of the combined object or its peaks seems
  to agree reasonably well with the BTF, but the low signal-to-noise
  nature of the HI observations makes further analysis or
  interpretation difficult.  We also caution that the WSRT
  observations do 
  not recover the total HI flux of ALFALFA (see Section
  \ref{AGC229383}), so 
  the HI masses of the two peaks do not sum to the total observed
  mass. 
}

We note that inclination effects could slightly modify our
determination of rotation velocities, but that the discrepancies from
the BTF are larger than can be accounted for by changing the
inclination angle of the objects. If
the distance to the HI1232+20 system was significantly smaller, then
they would be in better agreement with the BTF. In order for
AGC~229385 to agree with the BTF it would need to be at a distance of
$\sim$$4.4$~Mpc, and AGC~229384 would need to be at a distance of 
$\sim$$5.4$~Mpc. This smaller distance is unlikely given other
constraints (e.g., the lack of resolved stars in our optical
observations, see Section~\ref{denv}).

We also consider the ratio of dynamical mass to HI mass
($M_{\rm dyn} / M_{\rm HI}$) as another independent constraint on the
distance by assuming a typical value of 
$M_{\rm dyn} / M_{\rm HI}$$=$$10$. We measure an effective velocity
($v_{\rm eff}^2$$=$$ v_{\rm rot}^2 + 3\sigma_v^2$) for AGC~229385 of
$v_{\rm eff}$$=$$21.2$ km/s, 
so this relationship implies a distance of $4.3$~Mpc. If we assume no
dark matter (e.g., $M_{\rm dyn} / M_{\rm HI}$$=$$1.3$), we find a distance of
$32.8$~Mpc instead. The HI kinematics of AGC~229384 and AGC~229383 make it
difficult to measure $v_{\rm rot}$, which is required to determine
$v_{\rm eff}$.

\subsection{Formation scenarios}

Given the variety of observational constraints we have for the sources
in the HI1232+20 system, and the context from existing galaxy scaling
relations, we now consider some of the possible evolutionary scenarios
which could account for a system like this. 

{The blue color of the optical counterpart of AGC~229385
  suggests that it 
  may consist of a mostly young stellar population. If this object has
  only just begun forming stars, it may not have had enough time yet
  to convert a significant amount of its gas into stars, as reflected
  by the high gas mass-to-light ratio. However, the lack of H$\alpha$
  detection and the weak UV SFR ($\sim$$0.004 M_\odot/$yr) seem to
  indicate that it is only slowly forming stars.
  At this
  rate it would take more 
  than a Hubble time to generate enough mass in stars to return it to
  the normal relationship between HI and stellar mass (e.g., Figure
  \ref{huang}a).
}

It is difficult to find a {single} convincing explanation for
why one HI cloud has {an observable stellar population while
  its two nearby neighbors do not. } Obviously, this creates a
``fine-tuning'' problem of sorts. {Interactions between the
  members of this system may have triggered star formation in the most
  massive object. It is also possible that the system may have been
  perturbed by 
  an external object. Our analysis in Section \ref{isolation} showed
  that there was only one plausible perturber nearby, even given our
  generous assumptions about relative velocities and timescales for
  interactions.}

Alternatively, \citet{verde02} has suggested that low-mass
dark matter halos can contain neutral gas without ever forming stars,
under certain conditions. However, the members of this system have
substantial HI masses and are likely different from the low-mass dark
matter halos of \citet{verde02}. It may be that star formation has
only occurred in the HI cloud which is dense enough at its center to
exceed a {gas} density threshold (e.g.,
\citet{kennicutt98}). {Indeed, in studies of the outer disks of
  larger galaxies, star formation appears to cease below an HI surface
  density of $\sim1M_\odot/pc^2$ \citep{radburnsmith12, hunter13}. 
The HI distributions of the sources in the HI1232+20 system are mostly
below this surface density threshold, and only the peak of AGC~229385
substantially exceeds it. }

{The reasons behind the low gas surface density and
  over-extended HI distributions of these objects are not clear. In
  the case of AGC~229385} it is possible that feedback from the star 
formation (perhaps stronger in the past) has injected kinetic energy
into the HI, and temporarily expanded it. Or, we may be seeing the
results of a recent infall of cold gas into this system. However,
these objects have significant amounts of HI, and it is difficult to
justify an inflow scenario which could provide enough gas
{while maintaining a low gas density and inhibit star
  formation.}

It is also possible that these objects may be the most massive knots
{in a} system of smaller HI clouds which are in the {midst
  of a tidal interaction or merger event.} AGC~229385 and
AGC~229384 may be the brightest HI 
peaks, and could be the products of {recent} mergers of smaller HI
clumps. The extended nature of AGC~229383 and {its two HI peaks} 
are suggestive of an extended HI
distribution with local peaks and clumps. However, the ALFALFA
observations are very sensitive to faint HI emission, and given the
good agreement between WSRT and ALFALFA HI fluxes for AGC~229385 and
AGC~229384, it seems unlikely that there is a substantial amount of
unseen extended gas {in most of the system}.

{The HI1232+20 system does not seem to be a tidal feature or
  remnant of a larger object, based on the lack of a clear connection
  to any possible external perturbing object. However, tidal
  interactions between members of the system may be important to their
  individual star formation histories. The HI clouds in this system
  have extremely low star formation efficiencies and low gas
  densities. These low gas densities are especially difficult to
  reconcile with the significant HI masses of these sources. The HI mass of
  AGC~229385 is greater than that of the LMC, and AGC~229383 has
  more HI than some nearby star-forming dwarf irregular galaxies. Even
  if recent tidal interactions between the sources may have triggered
  a burst of star formation in AGC~229385, the overall properties of
  the HI1232+20 system are 
  difficult to convincingly interpret and explain.
}



\section{Summary}
\label{summary}

In this work we present the discovery of the HI1232+20 system of HI
sources, 
drawn from the 
 {sample of (almost) dark extragalactic sources in ALFALFA.}
This system
defies conventional explanations and our HI synthesis imaging and deep
optical observations have revealed a set of objects with properties
that are difficult to reconcile with typical scaling relations. The
most massive of its members (AGC~229385, 
$M_{HI}$$=$$7.2$$\times$$10^8 M_\odot$) has a  
weak stellar counterpart, detected in UV and ultra-deep optical
imaging, with a peak surface brightness of
$\mu_{g'}$$=$$26.4$~\magsec. It has the most extreme well-measured gas
mass-to-light ratio 
in the literature ($M_{HI}/L_B = 38$), and its absolute magnitude is
only $M_{g'} = -12.9$~mag ($M_\star = 1.5$$\times$$10^6 M_\odot$),
{assuming a distance of $25$~Mpc}. We do not
detect optical counterparts for 
the other two members, but place upper limits on their
absolute magnitudes of $M_{g'} > -11.3$~mag ($M_\star < 3$$\times$$10^5
M_\odot$). The HI kinematics of the three objects in this system are
inconsistent with typical galaxy scaling relations, with HI
distributions that are too 
extended and too slowly rotating for their HI mass. This group appears
on the sky just outside of the projected virial radius of the Virgo
Cluster, but is otherwise isolated from any nearby galaxies.

The HI1232+20 system is difficult to explain completely,
but may be an example of objects just above and just below a threshold
for star formation. The most massive of the three sources is forming
stars, but may have only recently started to do so. The other two
sources have no observational signatures of star formation, so there
may be 
some mechanism inhibiting this process. { Sources like these
  are very rare in the ALFALFA survey, especially at such large HI
  masses. }
As observations of the HI1232+20 system
continue we hope to learn more about its history and role in galaxy
formation and evolution.


\section{Acknowledgments}

We thank the entire ALFALFA team for their efforts in observing
   and data processing that produced the ALFALFA source catalog.
We also thank S. Huang for providing the data shown in Figure
\ref{huang}, from \citet{huang12}.
{We thank the anonymous referee for very helpful feedback and
  comments which have improved this work.}
  The ALFALFA work at Cornell is supported by NSF grants
  AST-0607007 and AST-1107390 to R.G. and M.P.H. and by grants
  from the Brinson Foundation.
J.M.C. is supported by NSF grant 1211683.
K.L.R. \& W.F.J. are supported by NSF Early Career Development Award
AST-0847109.
This research has made extensive use of the invaluable NASA/IPAC
Extragalactic Database 
(NED) which is operated by the Jet Propulsion Laboratory, California
Institute of Technology, under contract with the National Aeronautics
and Space Administration. 
This research has also made use of NASA's Astrophysics Data System.


\begin{thebibliography}{}




\bibitem[Ahn et al.(2012)]{ahn12} Ahn, C.~P., Alexandroff, 
R., Allende Prieto, C., et al.\ 2012, \apjs, 203, 21 



\bibitem[Bellazzini et al.(2015)]{Bellazzini15}	Bellazzini, M.,
  Magrini, L., Mucciarelli, A., et al.\ 2015, \apjl \, (in press)
  (arXiv:1501.06305) 



\bibitem[Belokurov et al.(2007)]{2007ApJ...654..897B} Belokurov, V., 
Zucker, D.~B., Evans, N.~W., et al.\ 2007, \apj, 654, 897 


\bibitem[Bernstein-Cooper et al.(2014)]{berncoop14} 
Bernstein-Cooper, E.~Z., Cannon, J.~M., Elson, E.~C., et al.\ 2014, \aj, 
148, 35 


\bibitem[Bianchi et al.(2014)]{bianchi14} Bianchi, L., Conti, A., 
\& Shiao, B.\ 2014, Advances in Space Research, 53, 900 


\bibitem[Binggeli et al.(1985)]{binggeli85} Binggeli, B., Sandage, 
A., \& Tammann, G.~A.\ 1985, \aj, 90, 1681 


\bibitem[Bressan et al.(2012)]{bressan12} Bressan, A., Marigo, 
P., Girardi, L., et al.\ 2012, \mnras, 427, 127 




\bibitem[Brinchmann et al.(2004)]{brinchmann04} Brinchmann, J., 
Charlot, S., White, S.~D.~M., et al.\ 2004, \mnras, 351, 1151 




\bibitem[Broeils \& Rhee(1997)]{BR97} Broeils, A.~H., \& Rhee, M.-H.\ 1997, \aap, 324, 877 


\bibitem[Cannon et al.(2011)]{cannon11} Cannon, J.~M., 
Giovanelli, R., Haynes, M.~P., et al.\ 2011, \apjl, 739, LL22 

\bibitem[Cannon et al.(2015)]{cannon15} Cannon, J.~M., 
Martinkus, C.~P., Leisman, L., et al.\ 2015, \aj, 149, 72 


\bibitem[Chengalur et al.(1995)]{chengalur95} Chengalur, J.~N., 
Giovanelli, R., \& Haynes, M.~P.\ 1995, \aj, 109, 2415 


\bibitem[Davies et al.(2004)]{davies04} Davies, J., Minchin, R., 
Sabatini, S., et al.\ 2004, \mnras, 349, 922 


\bibitem[Davies et al.(2006)]{davies06} Davies, J.~I., Disney, 
M.~J., Minchin, R.~F., Auld, R., \& Smith, R.\ 2006, \mnras, 368, 1479 


\bibitem[de Blok \& Walter(2000)]{deblok2000} de Blok, W.~J.~G., \& Walter, F.\ 2000, Mapping the Hidden Universe: The Universe behind the Milky Way - The Universe in HI, 218, 357 

\bibitem[Disney(1976)]{disney76} Disney, M.~J.\ 1976, \nat, 263, 
573 




\bibitem[Doi et al.(2010)]{doi10} Doi, M., Tanaka, M., 
Fukugita, M., et al.\ 2010, \aj, 139, 1628 


\bibitem[Doyle et al.(2005)]{doyle05} Doyle, M.~T., Drinkwater, 
M.~J., Rohde, D.~J., et al.\ 2005, \mnras, 361, 34 


\bibitem[Duc \& Bournaud(2008)]{duc08} Duc, P.-A., \& Bournaud, F.\ 2008, \apj, 673, 787 


\bibitem[Dutton et al.(2007)]{dutton07} Dutton, A.~A., van den 
Bosch, F.~C., Dekel, A., \& Courteau, S.\ 2007, \apj, 654, 27 


\bibitem[van Eymeren et al.(2009)]{vaneymeren2009} van Eymeren, J., Marcelin, M., Koribalski, B.~S., et al.\ 2009, \aap, 505, 105 



\bibitem[Faber \& Jackson(1976)]{fj76} Faber, S.~M., \& Jackson, R.~E.\ 1976, \apj, 204, 668 


\bibitem[Falco et al.(1999)]{falco99} Falco, E.~E., Kurtz, 
M.~J., Geller, M.~J., et al.\ 1999, \pasp, 111, 438 


\bibitem[Ferrarese et al.(2012)]{ferrarese12} Ferrarese, L., 
C{\^o}t{\'e}, P., Cuillandre, J.-C., et al.\ 2012, \apjs, 200, 4 


\bibitem[Geha et al.(2006)]{geha06} Geha, M., Blanton, M.~R., 
Masjedi, M., \& West, A.~A.\ 2006, \apj, 653, 240 




\bibitem[Giovanelli et al.(2005)]{giovanelli05} Giovanelli, R., 
Haynes, M.~P., Kent, B.~R., et al.\ 2005, \aj, 130, 2598 


\bibitem[Giovanelli et al.(2013)]{giovanelli13} Giovanelli, R., 
Haynes, M.~P., Adams, E.~A.~K., et al.\ 2013, \aj, 146, 15 


\bibitem[Gopu et al.(2014)]{gopu14} Gopu, A., Hayashi, S., 
\& Young, M.~D.\ 2014, Astronomical Data Analysis Software and Systems XXIII, 485, 417 


\bibitem[Gratier et al.(2010)]{gratier10} Gratier, P., Braine, J., Rodriguez-Fernandez, N.~J., et al.\ 2010, \aap, 522, AA3 

\bibitem[Gunn et al.(1998)]{gunn98} Gunn, J.~E., Carr, M., 
Rockosi, C., et al.\ 1998, \aj, 116, 3040 


\bibitem[Hao et al.(2011)]{hao11} Hao, C.-N., Kennicutt, 
R.~C., Johnson, B.~D., et al.\ 2011, \apj, 741, 124 




\bibitem[Haynes et al.(2011)]{haynes11} Haynes, M.~P., 
Giovanelli, R., Martin, A.~M., et al.\ 2011, \aj, 142, 170 


\bibitem[Hopkins \& Beacom(2006)]{hopkins06} Hopkins, A.~M., \& Beacom, J.~F.\ 2006, \apj, 651, 142 


\bibitem[Huang et al.(2012a)]{huang12dwarfs} Huang, S., Haynes, M.~P., 
Giovanelli, R., et al.\ 2012a, \aj, 143, 133  


\bibitem[Huang et al.(2012b)]{huang12} Huang, S., Haynes, M.~P., 
Giovanelli, R., \& Brinchmann, J.\ 2012b, \apj, 756, 113   


\bibitem[Hunter et al.(2013)]{hunter13} Hunter, D.~A., 
Elmegreen, B.~G., Rubin, V.~C., et al.\ 2013, \aj, 146, 92 




\bibitem[Jester et al.(2005)]{jester05} Jester, S., Schneider, 
D.~P., Richards, G.~T., et al.\ 2005, \aj, 130, 873 


 \bibitem[Jones et al.(2015)]{in preparation}  Jones, M., \etal in
   preparation


\bibitem[Karachentsev \& Nasonova(2010)]{karnas10} Karachentsev, I.~D., \& Nasonova, O.~G.\ 2010, \mnras, 405, 1075 


\bibitem[Karachentsev et al.(2011)]{kar11} Karachentsev, 
I.~D., Nasonova, O.~G., \& Courtois, H.~M.\ 2011, \apj, 743, 123 


\bibitem[Karachentsev et al.(2014)]{kar14} Karachentsev, 
I.~D., Tully, R.~B., Wu, P.-F., Shaya, E.~J., 
\& Dolphin, A.~E.\ 2014, \apj, 782, 4 


\bibitem[Kennicutt(1998)]{kennicutt98} Kennicutt, R.~C., Jr.\ 1998, 
\apj, 498, 541 


\bibitem[Kim et al.(1998)]{kim98} Kim, S., Staveley-Smith, L., Dopita,   M.~A., et al.\ 1998, \apj, 503, 674 


\bibitem[Kotulla(2014)]{kotulla14} Kotulla, R.\ 2014, 
Astronomical Data Analysis Software and Systems XXIII, 485, 375 


\bibitem[Lee et al.(2003)]{lee03} Lee, H., McCall, M.~L., 
Kingsburgh, R.~L., Ross, R., \& Stevenson, C.~C.\ 2003, \aj, 125, 146 


\bibitem[Lelli et al.(2010)]{lelli2010} Lelli, F., Fraternali, F., \& Sancisi, R.\ 2010, \aap, 516, AA11 

\bibitem[Lucy(1974)]{lucy74} Lucy, L.~B.\ 1974, \aj, 79, 745 


\bibitem[Madau et al.(1998)]{madau98} Madau, P., Pozzetti, L., 
\& Dickinson, M.\ 1998, \apj, 498, 106 


\bibitem[Martin et al.(2005)]{martin05} Martin, D.~C., Fanson, 
J., Schiminovich, D., et al.\ 2005, \apjl, 619, L1 


\bibitem[Martin et al.(2009)]{martin09} Martin, A.~M., 
Giovanelli, R., Haynes, M.~P., et al.\ 2009, \apjs, 183, 214 


\bibitem[Martin et al.(2010)]{martin10} Martin, A.~M., 
Papastergis, E., Giovanelli, R., et al.\ 2010, \apj, 723, 1359 



\bibitem[Masters(2005)]{masters05} Masters, K.~L.\ 2005, 
Ph.D.~Thesis,  


\bibitem[Matsuoka et al.(2012)]{matsuoka12} Matsuoka, Y., Ienaka, N., Oyabu, S., Wada, K., \& Takino, S.\ 2012, \aj, 144, 159 


\bibitem[Matthews et al.(1998)]{matthews98} Matthews, L.~D., van 
Driel, W., \& Gallagher, J.~S., III 1998, \aj, 116, 2196 


\bibitem[McGaugh et al.(2000)]{mcgaugh2000} McGaugh, S.~S., 
Schombert, J.~M., Bothun, G.~D., 
\& de Blok, W.~J.~G.\ 2000, \apjl, 533, L99 


\bibitem[McGaugh(2005)]{mcgaugh05} McGaugh, S.~S.\ 2005, \apj, 
632, 859 

\bibitem[McGaugh(2012)]{mcgaugh12} McGaugh, S.~S.\ 2012, \aj, 
143, 40 

\bibitem[McGaugh \& de Blok(1997)]{mcgaugh97} McGaugh, S.~S., \& de Blok, W.~J.~G.\ 1997, \apj, 481, 689 

\bibitem[McGaugh \& Schombert(2014)]{mcgsch14} McGaugh, S.~S., \& Schombert, J.~M.\ 2014, \aj, 148, 77 




\bibitem[Meyer et al.(2004)]{meyer04} Meyer, M.~J., Zwaan, 
M.~A., Webster, R.~L., et al.\ 2004, \mnras, 350, 1195 








\bibitem[Morrissey et al.(2007)]{morrissey07} Morrissey, P., Conrow, T., Barlow, T.~A., et al.\ 2007, \apjs, 173, 682 


\bibitem[Murphy et al.(2011)]{murphy11} Murphy, E.~J., Condon, 
J.~J., Schinnerer, E., et al.\ 2011, \apj, 737, 67 


\bibitem[Nidever et al.(2013)]{nidever13} Nidever, D.~L., Ashley, T., Slater, C.~T., et al.\ 2013, \apjl, 779, LL15 



\bibitem[Planck Collaboration et al.(2014)]{Planck14} Planck Collaboration, Ade, P.~A.~R., Aghanim, N., et al.\ 2014, \aap, 566, AA54 


\bibitem[Radburn-Smith et al.(2012)]{radburnsmith12} Radburn-Smith, 
D.~J., Ro{\v s}kar, R., Debattista, V.~P., et al.\ 2012, \apj, 753, 138 



\bibitem[Rhode et al.(2013)]{rhode13} Rhode, K.~L., Salzer, 
J.~J., Haurberg, N.~C., et al.\ 2013, \aj, 145, 149 


\bibitem[Roberts \& Haynes(1994)]{robertshaynes94} Roberts, M.~S., \& Haynes, M.~P.\ 1994, \araa, 32, 115 



\bibitem[Rosenberg \& Schneider(2003)]{rosenbergschneider03} Rosenberg, J.~L., \& Schneider, S.~E.\ 2003, \apj, 585, 256 


\bibitem[Rosenberg \& Schneider(2000)]{rosenbergschneider00} Rosenberg, J.~L., \& Schneider, S.~E.\ 2000, \apjs, 130, 177 


\bibitem[Rudick et al.(2010)]{rudick10} Rudick, C.~S., Mihos, 
J.~C., Harding, P., et al.\ 2010, \apj, 720, 569 


\bibitem[Saintonge(2007)]{saintonge07} Saintonge, A.\ 2007, \aj, 
133, 2087 


\bibitem[Salim et al.(2007)]{salim07} Salim, S., Rich, R.~M., 
Charlot, S., et al.\ 2007, \apjs, 173, 267 





\bibitem[Sancisi et al.(1990)]{sancisi90} Sancisi, R., Broeils, 
A., Kamphuis, J., 
\& van der Hulst, T.\ 1990, Dynamics and Interactions of Galaxies, 304 



\bibitem[Sandage(1976)]{sandage76} Sandage, A.\ 1976, \aj, 81, 
954 


\bibitem[Sault et al.(1995)]{sault95} Sault, R.~J., Teuben, 
P.~J., 
\& Wright, M.~C.~H.\ 1995, Astronomical Data Analysis Software and Systems IV, 77, 433 


\bibitem[Schlafly \& Finkbeiner(2011)]{sf11} Schlafly, E.~F., \& Finkbeiner, D.~P.\ 2011, \apj, 737, 103 


\bibitem[Schlegel et al.(1998)]{sfd98} Schlegel, D.~J., 
Finkbeiner, D.~P., \& Davis, M.\ 1998, \apj, 500, 525 



\bibitem[Schombert et al.(2011)]{schombert11} Schombert, J., 
Maciel, T., \& McGaugh, S.\ 2011, Advances in Astronomy, 2011, 143698 



\bibitem[Schombert \& McGaugh(2014)]{schombert14} Schombert, J., \& McGaugh, S.\ 2014, \pasa, 31, e036 


\bibitem[Serra et al.(2012)]{serra12} Serra, P., Oosterloo, T., 
Morganti, R., et al.\ 2012, \mnras, 422, 1835 


\bibitem[Stil \& Israel(2002)]{stilisrael02} Stil, J.~M., \& Israel, F.~P.\ 2002, \aap, 389, 29 


\bibitem[Swaters et al.(2002)]{swaters02} Swaters, R.~A., van Albada, T.~S., van der Hulst, J.~M., \& Sancisi, R.\ 2002, \aap, 390, 829 


\bibitem[Taylor \& Webster(2005)]{TaylorWebster2005} Taylor, E.~N., \& Webster, R.~L.\ 2005, \apj, 634, 1067 

\bibitem[Toribio et al.(2011)]{toribio11} Toribio, M.~C., 
Solanes, J.~M., Giovanelli, R., Haynes, M.~P., 
\& Martin, A.~M.\ 2011, \apj, 732, 93 


\bibitem[Tully \& Fisher(1977)]{tf77} Tully, R.~B., \& Fisher, J.~R.\ 1977, \aap, 54, 661 

\bibitem[Tully \& Fisher(1987)]{tf87} Tully, R.~B., \& Fisher, J.~R.\ 1987, Annales de Geophysique,  


\bibitem[van Zee et al.(1997)]{vanzee97} van Zee, L., Haynes, 
M.~P., Salzer, J.~J., \& Broeils, A.~H.\ 1997, \aj, 113, 1618 


\bibitem[Verde et al.(2002)]{verde02} Verde, L., Oh, S.~P., 
\& Jimenez, R.\ 2002, \mnras, 336, 541 



\bibitem[Wang et al.(2013)]{wang13} Wang, J., Kauffmann, G., 
J{\'o}zsa, G.~I.~G., et al.\ 2013, \mnras, 433, 270 


\bibitem[Warmels(1988)]{warmels88} Warmels, R.~H.\ 1988, \aaps, 72, 427 


\bibitem[Warren et al.(2007)]{warren07} Warren, B.~E., Jerjen, 
H., \& Koribalski, B.~S.\ 2007, \aj, 134, 1849 

\bibitem[Warren et al.(2004)]{warren04} Warren, B.~E., Jerjen, 
H., \& Koribalski, B.~S.\ 2004, \aj, 128, 1152 



\bibitem[Williams et al.(1996)]{williams96} Williams, R.~E., 
Blacker, B., Dickinson, M., et al.\ 1996, \aj, 112, 1335 



\bibitem[Witt et al.(2008)]{witt08} Witt, A.~N., Mandel, S., 
Sell, P.~H., Dixon, T., \& Vijh, U.~P.\ 2008, \apj, 679, 497 


\bibitem[Wright et al.(2010)]{wright10} Wright, E.~L., 
Eisenhardt, P.~R.~M., Mainzer, A.~K., et al.\ 2010, \aj, 140, 1868 


\bibitem[Young et al.(2013)]{young13} Young, M.~D., Gopu, A., 
Hayashi, S., 
\& Cox, J.~A.\ 2013, Astronomical Data Analysis Software and Systems XXII, 475, 337 



\end{thebibliography}
\end{document}